\documentclass[conference]{IEEEtran}
\IEEEoverridecommandlockouts

\usepackage{cite}
\usepackage{amsmath,amssymb,amsfonts}
\usepackage{algorithmic}
\usepackage{caption}
\usepackage{graphicx}
\usepackage{textcomp}
\usepackage{multicol}
\usepackage{multirow}
\usepackage{xcolor}
\usepackage[hyphens]{url}

\def\BibTeX{{\rm B\kern-.05em{\sc i\kern-.025em b}\kern-.08em
    T\kern-.1667em\lower.7ex\hbox{E}\kern-.125emX}}
\begin{document}

\pdfpagewidth=8.5in
\pdfpageheight=11in

\newcommand{\iscasubmissionnumber}{3554}

\pagenumbering{arabic}

\title{PC-Indexed Data Address Translation}
\author{Shyam Murthy, Gurindar S Sohi\\
University of Wisconsin-Madison\\
\{shyamm, sohi\}@cs.wisc.edu}

\maketitle
\thispagestyle{plain}
\pagestyle{plain}


\title{PC-Indexed Data Address Translation}


\maketitle 

\begin{abstract}

This paper proposes a novel way to assist conventional data address translation.  The approach, \textit{PC-Indexed Data Address Translation (PCAX)}, uses the PC of a load instruction, and not a data virtual address, to obtain the page table entry (PTE) for the data of the load instruction.  PCAX is intended to be used for a small subset of the static loads in a program.  We observe that: (i) a small subset of static loads is responsible for most of the misses in a data TLB (DTLB), and (ii) often a dynamic instance of a static load instruction accesses the same PTE as the last dynamic instance,
and consider PCAX for this subset.
With PCAX the effective miss rate of a conventional DTLB can be cut down by a factor of 2-3X, in many cases, and even more in some cases.  PCAX is also beneficial in reducing the number of secondary TLB (STLB) misses.
Since the PCAX tables are accessed alongside instruction fetch, they can tolerate relatively long access latencies while still frequently providing a valid PTE even \textit{before} data address calculation. This yields an average performance improvement of 1.7\% while reducing data address translation energy by 7\% across 84 server traces.

\end{abstract}

\section{Introduction}

Translation from virtual to physical memory addresses is a fundamental operation that is carried out on every data memory access and is critical to the efficient processing of memory load (and store) instructions.
A \textit{page table entry (PTE)} in a page table holds a translation from a virtual page number (VPN) to a physical page number (PPN), along with other information. 
\textit{Data Translation Lookaside Buffers (DTLBs)},
accessed using the data address (virtual page number) of a load (or store) instruction, 
are used to cache PTEs to provide fast access and improve address translation efficiency.
Given the criticality of data address translation to overall system performance,
a plethora of techniques spanning both hardware and (system) software,
have been proposed over the past several decades to improve its performance and efficiency.

Since the processing of load (and store) instructions is governed by the addresses of the data accessed, most techniques to improve their processing, including DTLBs for address translation, are based upon the empirical properties of the addresses of the data memory accesses.  
However, some techniques to optimize load instruction processing rely on the identity of the instruction, specifically the instruction's program counter (PC), rather than the data memory addresses. A prime example is memory dependence prediction, also referred to as memory disambiguation in some commercial processors. This technique is widely used in high-end processors to improve load scheduling by predicting dependencies between load and store instructions based on their PCs, without relying on the actual data memory addresses they access \cite{moshovos1997dynamic,chrysos1998memory}.
These pairs are tracked in a small PC-indexed table. This table is consulted to predict whether a load can be benefically executed before prior stores, without the determination of the data addresses of the load and store instructions.

In a similar vein, in this paper, we propose a non-traditional approach, \textit{PC-Indexed Data Address Translation (PCAX)}, to assist conventional data address translation which employs a \textit{conventional DTLB} (which we refer to as a \textit{CDTLB} in this paper).
The approach relies on using the identity of an instruction (the instruction PC), and not on the actual data address to obtain a PTE for the data address that is expected to be accessed by the instruction.  
PCAX is intended to be used for a small subset of load instructions: those whose data translation access is expected to miss in a CDTLB.

PCAX is based on two important and novel empirical observations.
First, \textit{a small percentage (and number) of static load instructions of a program account for a significant number of CDTLB misses}.
Second, \textit{quite often, when a dynamic instance of a static load has a CDTLB miss, it accesses the same virtual page (and the same PTE) as the previous dynamic instance}. 

These observations suggest that a (relatively small) table of PTEs, accessed with the PC of a load instruction, a \textit{PC-Indexed Data Address Translation Table (PCAT)}, could provide the PTEs that would normally be obtained as a result of a DTLB miss.  PCAX employs PCATs (loosely) coupled to the instruction caches.
A PCAT is selectively accessed during instruction fetch and provides, with a reasonable accuracy, a needed PTE \textit{before the data address is generated, the CDTLB accessed, and a miss occurs.}
A PCAT is not meant to replace a CDTLB but rather augment it, by providing the PTE for a reference that likely will miss in the CDTLB.

Empirical results show that PCAX can be extremely effective, reducing the effective CDTLB miss rate by 2-3X in most cases for the benchmarks considered.  PCAX is also useful in reducing the secondary TLB (STLB) misses. These miss reductions lead to improvements in both performance and energy efficiency in most cases.

\section{Motivation}
\label{sec:Motivation}


A \textit{static load} is a load at a given PC, a \textit{dynamic load} is a dynamic instance of a given static load. A \textit{problem load} is a static load whose dynamic instances result in CDTLB misses.


We present empirical observations to motivate PC-indexed address translation. These observations focus on the characteristics of misses in conventional DTLBs (CDTLBs), when the load data address is submitted to the CDTLB.
For benchmarks, we consider a set of 84 traces from Qualcomm, for the ARM architecture.
These traces, details of which are presented in Section \ref{sec:Eval}, have been used
in several recent papers \cite{asheim2023storage,chacon2023characterization,ros2021cost}. 
We consider a single page size of 4KB, and the miss rates, with a 64-entry CDTLB, range from 1\% to 31\%,
and the MPKIs from 1 to 53, as we shall see in section \ref{sec:BenchChars}.

\begin{figure}[t]
  \includegraphics[width=1.0\linewidth]{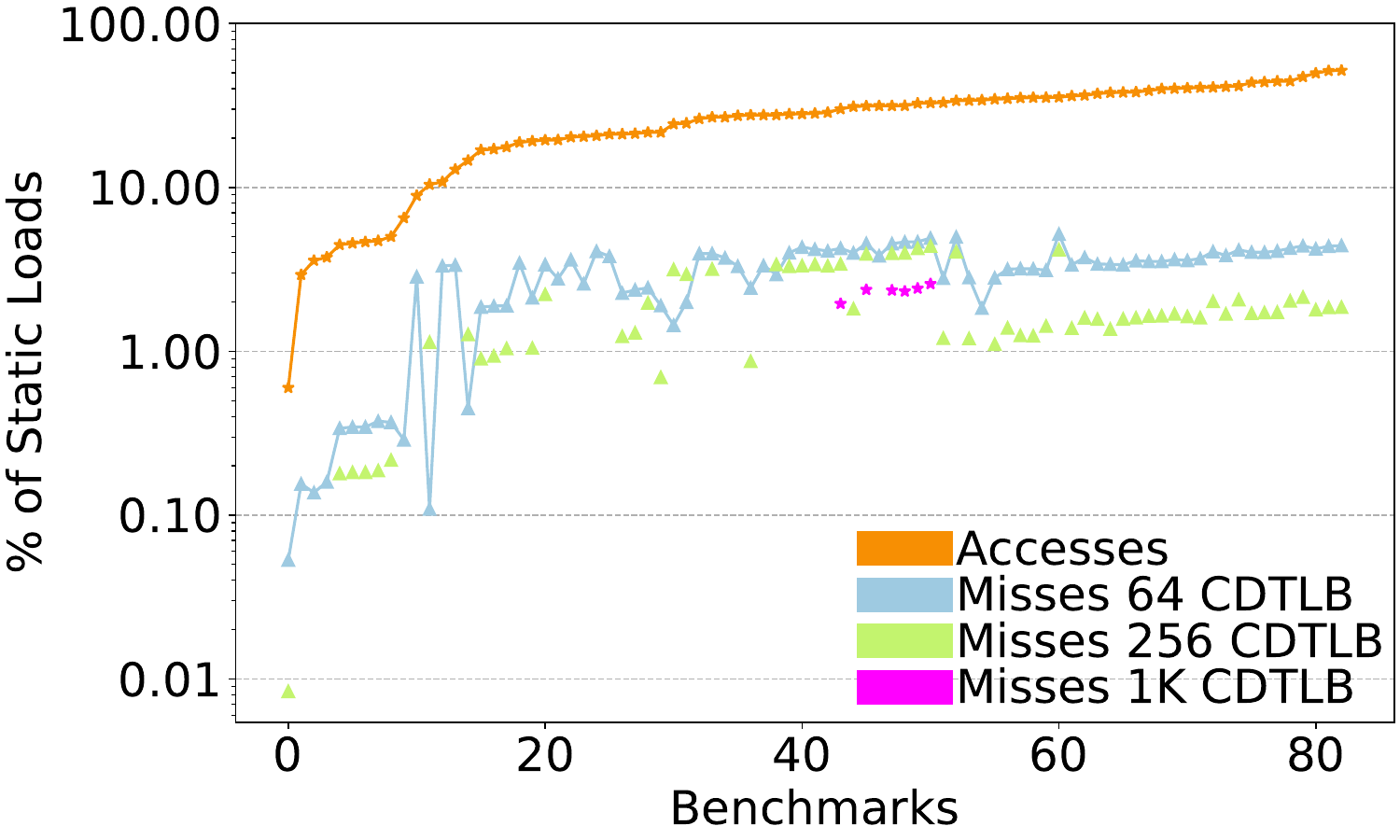}
	\caption {Percentage of static loads that are problem loads}
  \label{fig:per-static-load-server-64-256-1k}

\end{figure}

Figure \ref{fig:per-static-load-server-64-256-1k} shows the percentage of static loads that account for 90\% of the: (i) accesses, and (ii) misses in a CDTLB. 
In the figure, the X-axis is the different benchmarks, and the Y-axis (note the log scale) is the percentage of static loads. We consider 3 CDTLB sizes: 64, 256 and 1024 entries, with 4-way associativity in each case.  The larger sizes are representative of L2 or \textit{secondary} TLBs.   
For a given CDTLB size, we only plot the cases for which the miss rate is greater than 1\%. By \textit{vast majority} we mean 90\% (of the total accesses or misses).

The observation we make from the figure, which we do not believe has been documented before, is that \textit{a very small percentage of static loads are problem loads accounting for a vast majority of the misses in the CDTLB}.
A related, known observation that is that a small percentage of static loads account for a large percentage of data cache misses \cite{abraham1993predictability,tyson1997managing}.
Similarly, \cite{vavouliotis2021morrigan} observes that a few instruction accesses account for a vast majority of the instruction TLB misses.
This percentage of problem loads is typically an order of magnitude smaller than the percentage of static loads accounting for the references. For example, with a 64-entry CDTLB, in the median case, about 30\% of the static loads account for the vast majority of CDTLB accesses, but only about 4\% of the static loads account for the vast majority of CDTLB misses. From  Figure \ref{fig:per-static-load-server-64-256-1k}
we also see that as the CDTLB size increases the percentage of static loads that are problem loads decreases.
In rare cases the number of problem loads is higher for bigger CDTLBs.  We address this in section \ref{sec:BenchChars}.
Overall, the percentage of problem loads is in the single digits.

\begin{figure}[t]
  \includegraphics[width=0.9\linewidth]{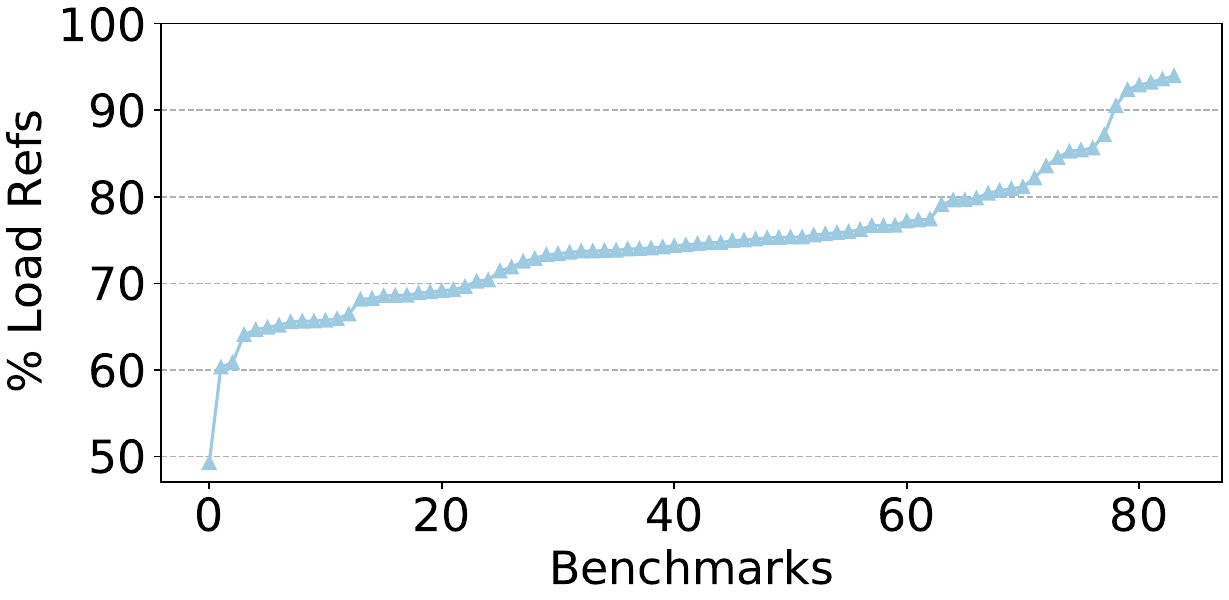}
	\caption {Loads Accessing same PTE}
  \label{fig:load-ref-to-prevpte}
\end{figure}

Figure \ref{fig:load-ref-to-prevpte} presents the percentage of all loads (both hits and misses in the CDLTB) that access the same PTE as the previous dynamic instance of the load. Observe that the percentage of loads accessing the same PTE as the previous instance is as high as 70-90\% for many of the benchmarks.  This result should not be surprising as many dynamic instances of a static load access proximate elements of a data structure, which are likely to be in proximate memory addresses, and thus mostly on the same page.

\begin{figure}[t]
  \includegraphics[width=1.0\linewidth]{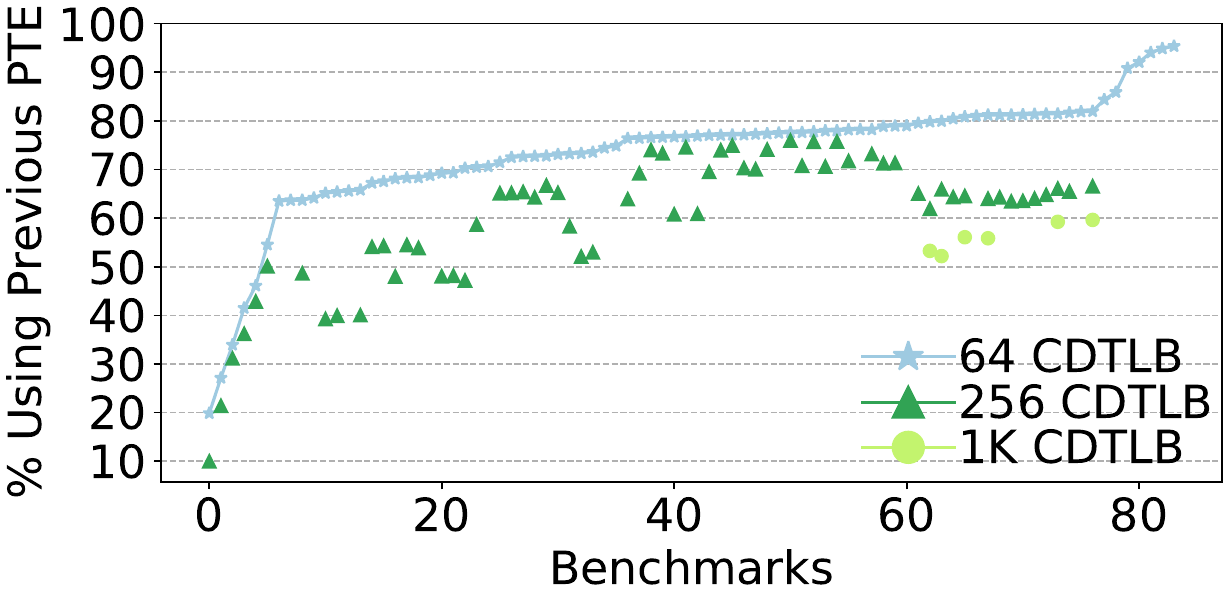}
	\caption {Problem loads accessing same PTE}
  \label{fig:prev-page-accesses-misses-server-64-256-1k-tlb}

\end{figure}

Figure \ref{fig:prev-page-accesses-misses-server-64-256-1k-tlb} presents the percentage of time that a dynamic instance of a static load identified as a problem load  accesses and misses on the same PTE as the last missing dynamic instance of the static (problem) load. The figure has 3 different DTLB sizes, 64, 256, and 1K entries. Observe that for a very large percentage of the problem loads, the PTE that is accessed on a miss is the same as the PTE for the prior instance of the problem load.
For example, at a median point, with a 64-entry CDTLB, about 75\% of the loads that miss in the CDTLB access the same  PTE as the prior dynamic instance. This percentage decreases as the CDTLB size increases. But even for the large CDTLB sizes, more than half of the problem loads access the same PTE as the prior dynamic instance.




\begin{figure}[t]
  \includegraphics[width=0.9\linewidth]{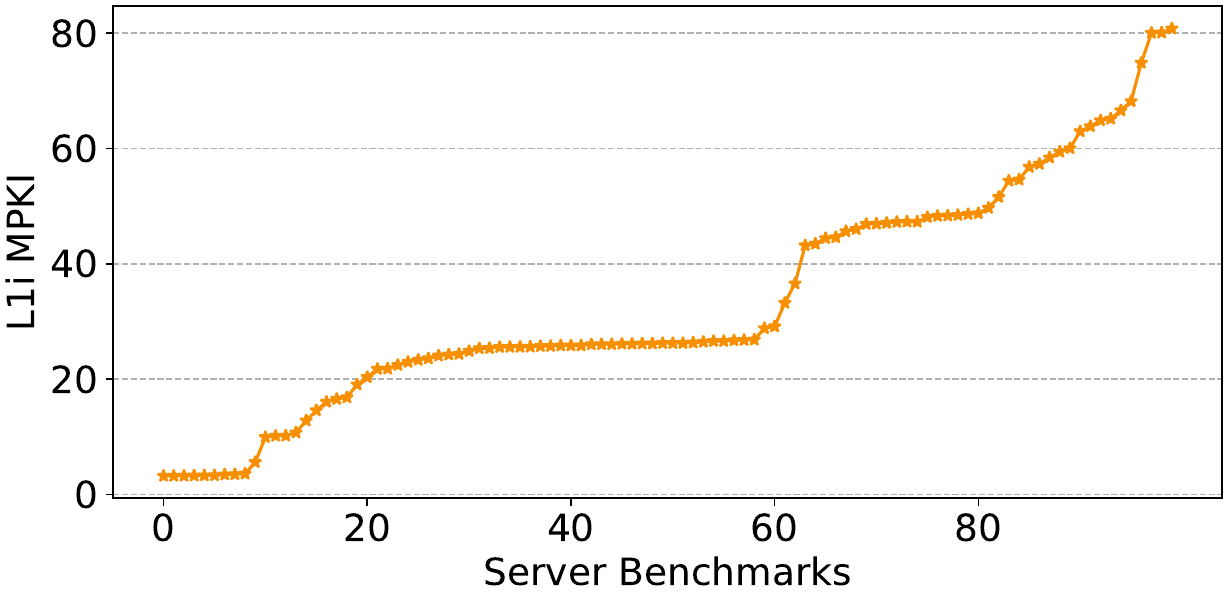}
	\caption {L1i MPKI}
  \label{fig:l1i-mpki}
\end{figure}

The empirical observations of Figures \ref{fig:per-static-load-server-64-256-1k} and
\ref{fig:prev-page-accesses-misses-server-64-256-1k-tlb}
suggest that if we were to track the last PTE for static loads that miss in the CDTLB,
in a table indexed by the load PC,
we could get the correct PTE for the data that a load instruction would reference,
\textit{and would likely miss in the CDTLB}, a fairly large percentage of the time.
Further, using the load PC, this correct (and CDTLB-missing) PTE could be obtained while the load was being fetched,
i.e., even before it is decoded and well before the address of the data it will reference is determined.
This tracking needs to be done only for a small percentage of the static loads in a program (we will quantify the absolute number of problem later in section \ref{sec:BenchChars}).


\begin{figure}[t]
  \includegraphics[width=0.9\linewidth]{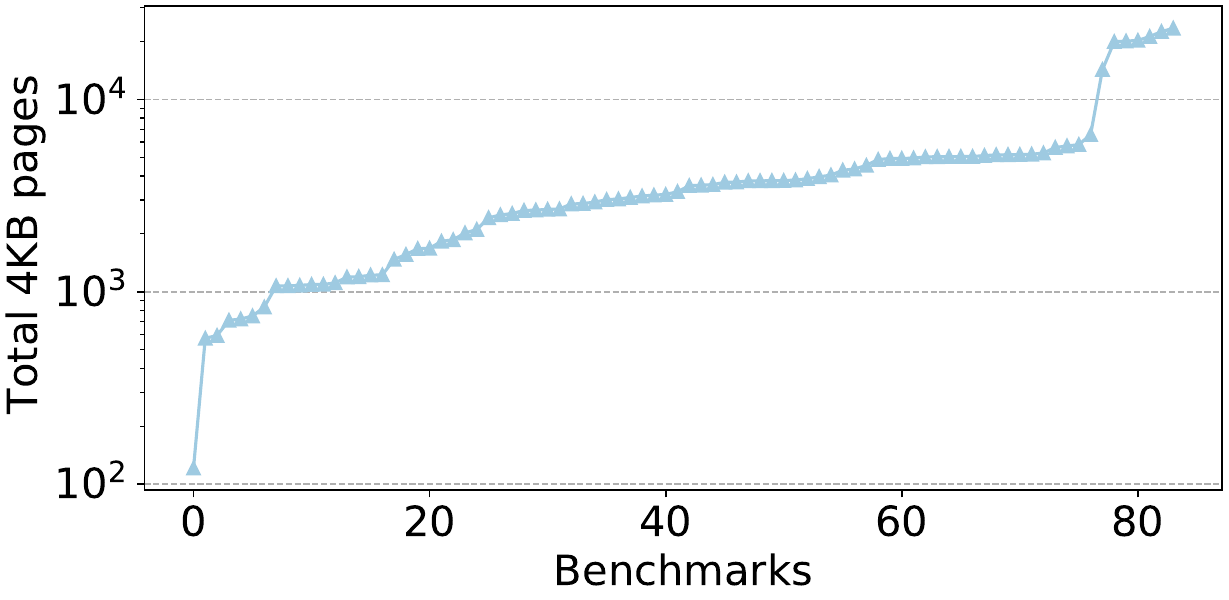}
	\caption {Total Data Pages Touched}
  \label{fig:tot-dpages-touched}
\end{figure}

Next we consider some key data page reference characteristics to motivate how the above observations could be exploited. Figure \ref{fig:tot-dpages-touched}
presents the number of 4KB data pages touched by the different applications (note log scale on Y-axis).  The data footprint is not very large.  Figure \ref{fig:l1i-mpki} presents the L1i MPKI touched by the difference benchmarks for an L1 instruction cache (L1i) of size 32 KB. As we see, the benchmarks also have large code footprints \cite{asheim2023storage,chacon2023characterization,ros2021cost};
for an L1 instruction cache (L1i) of size 32KB, there are 45 benchmarks with an L1i MPKI between 0-30, 11 benchmarks with an L1i MPKI between 30-45, 21 benchmarks with an L1i MPKI between 45-60 and 7 benchmarks with an L1i MPKI $>$ 60.
The large code footprint results in very frequent code movement from an L2 cache to the L1i instruction cache, 
and coupled with the data referencing patterns, which we see next 
collectively results in a significant number of CDTLB misses, as we shall see in section \ref{sec:BenchChars}.


\begin{figure}[t]
  \includegraphics[width=0.9\linewidth]{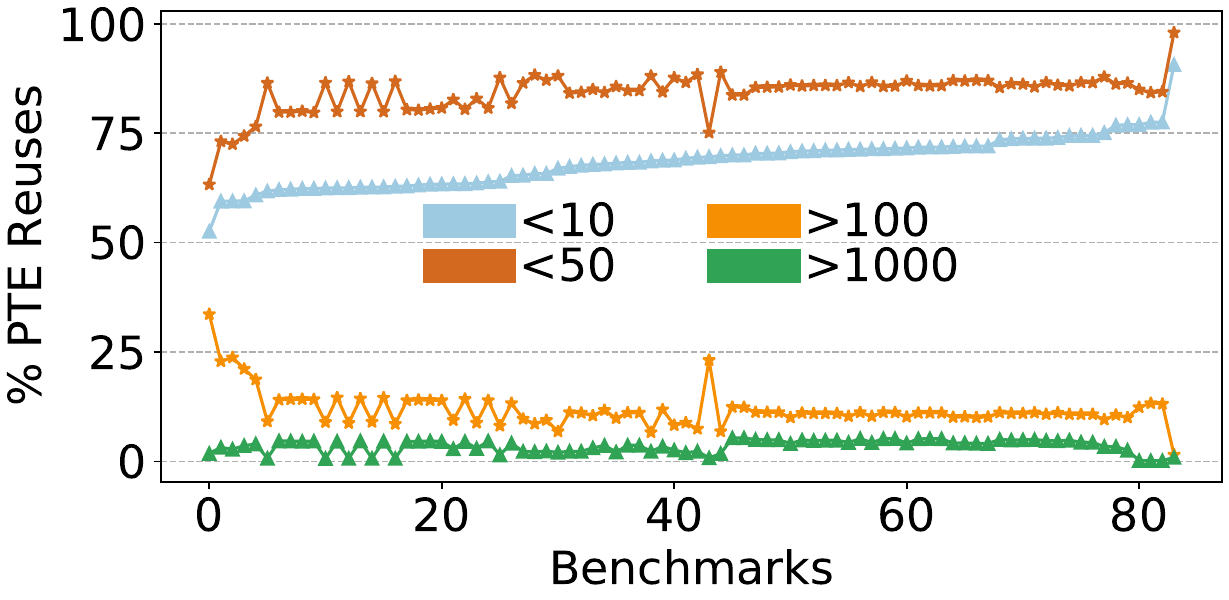}
	\caption {PTE Reuse Characteristics}
  \label{fig:pte-ref-char}
\end{figure}

\begin{figure}[t]
  \includegraphics[width=0.9\linewidth]{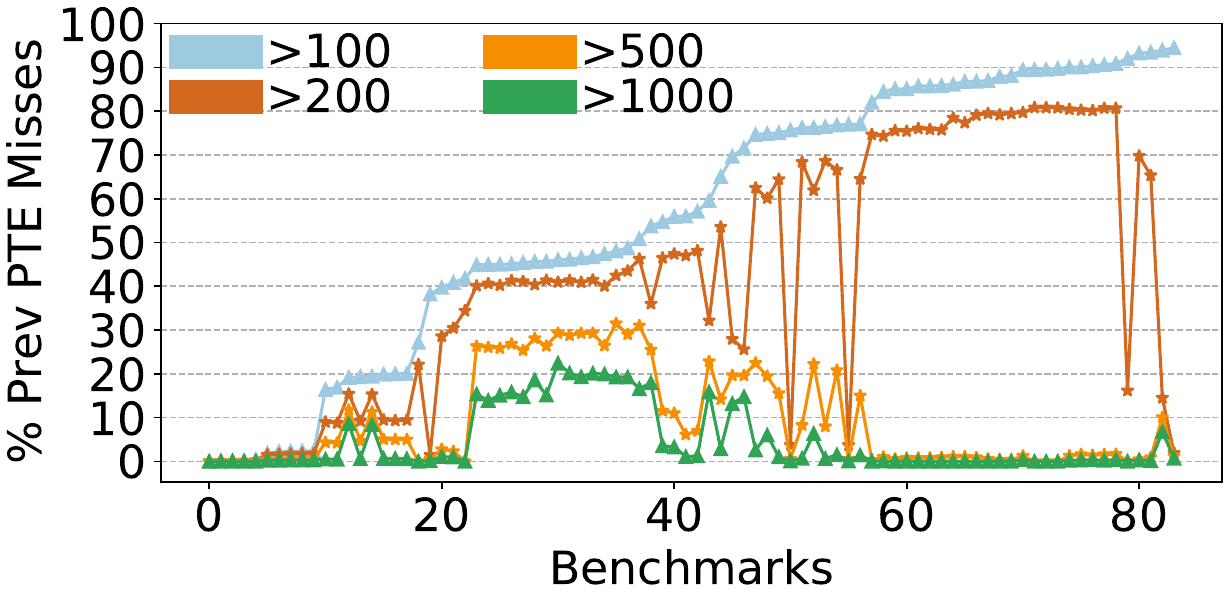}
	\caption {PTE Rereference Distance}
  \label{fig:reuse-dist}
\end{figure}


Figure \ref{fig:pte-ref-char} presents the reuse characteristics of PTEs, that is, after an initial load reference that brings a PTE into the CDTLB, what percentage of
reuses of the PTE (either from another dynamic instance of the same static load, or from dynamic instances of other static loads) occur within $<$ 10, $<$ 50, $>$ 100, or $>$ 1000
dynamic load instructions.  From the figure we see that a very large percentage of references (and reuses) to a PTE occur within fewer than 50 dynamic load references 
and a very small percentage are for $>$ 100 references.  Further, we have observed, data not shown, that dynamic instances of an average of 2-8 other static loads reference a PTE brought into a CDTLB by a dynamic instance of a static load.  These results should also not be surprising because we expect references to different fields of an element of a data structure (from different static loads), which are likely to be on the same page, to occur close to each other.


Figure \ref{fig:reuse-dist} presents the re-reference distance for a PTE. That is, between the time a PTE is brought into the CDTLB by a dynamic instance of a static load, how many other PTEs are referenced, on average, until the same PTE is re-referenced and brought into the CDTLB again by another dynamic instance of the same static load. Data is presented for re-reference distances of greater than 100, 200, 500, and 1000 PTEs. Many benchmarks have more than 50-60\% of such missing PTE references with a re-reference distance greater than 100 and 200.  Even the percentage of such misses with re-reference distances greater than 500 and 1000 is non-trivial (20-30\%) for some benchmarks. Such large re-reference distances lead to higher miss rates with relatively small-sized CDTLBs.

To put it together, the server workloads we study exhibit large instruction working sets that cause frequent eviction and later re-fetch a static load instruction from the L1i as execution moves across different code regions. As a result, successive dynamic instances of the same static load may be separated by substantial intervals in the dynamic instruction stream. During these intervals, references to other data pages displace the earlier PTE from the CDTLB. When the static load is eventually re-fetched into the L1i, it often accesses the same virtual page as before, yet the translation is no longer present in the CDTLB. It is the interaction of (i) large code footprints, which increase the temporal separation between dynamic instances of a load, and (ii) structured data referencing patterns, which preserve per-PC PTE reuse.

This is what provides the basis for the scheme we present next: associate PTEs with a small subset of loads (which miss in the CDTLB) and bring a PTE associated with a load when the load is brought back into the L1i so that it can be used when the load executes without incurring a CDTLB miss.   In other words, view the issue of CDTLB misses not from the viewpoint of the data reference stream, but from the viewpoint of the instruction reference stream which creates the data references that result in the CDTLB misses.

\section{PC-Indexed Address Translation}
\label{sec:Mechanism}














\subsection {Overview}

Motivated by the empirical observations of Section \ref{sec:Motivation}, we propose
\textit{PC-Indexed Address Translation (PCAX)}.
The idea is to maintain PTEs for a select group of load instructions in a
\textit{PC-Indexed Address Translation Table (PCAT)}.
When one of these select loads is fetched, the PCAT is accessed to find a corresponding entry.
The PTE from this entry is then provided in the expectation that it will be
the PTE that will be needed when the address of the data referenced by that load is
determined. The purpose of PCAX is to augment a CDTLB.
For most loads, the CDTLB provides the PTE for the accessed data.
For select load instructions, a (hopefully good) translation is provided by a PCAT.

\begin{figure}[t]
  \includegraphics[width=1.0\linewidth]{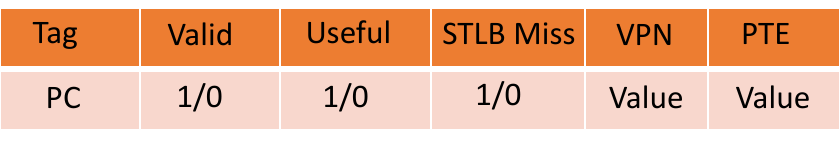}
	\caption {PCAT Tag and Data Entry}
  \label{fig:pcat-entry}
  \vspace{-10pt}
\end{figure}

\subsection {PCAT Structure and Operation}
\label{sec:pcat1-struc-op}


A PCAT is a cache with a tag array and a data array. Each entry, shown in Figure \ref{fig:pcat-entry}, contains: (i) tag bits, (ii) a virtual page number (VPN), 
(iii) the corresponding page table entry (PTE)—collectively referred to as the VPN-PTE pair, and 
(iv) two additional bits used for replacement decisions: a \textit{Useful} bit, and a \textit{Secondary TLB (STLB) Miss}  bit. The \textit{STLB Miss (STLBM)} bit is set when the entry is created as a result of a miss in the secondary TLB, and it is later used to guide replacement decisions.  
Each entry also includes a Valid bit. The tag and VPN-PTE fields are self-explanatory. We describe the usage of these bits in more detail later.

Since access to a PCAT entry is closely tied to an instruction access---a PCAT entry will not be accessed if the corresponding load instruction isn't also accessed---it is natural for the PCAT structures to mirror
the cache structures holding instructions. 
Accordingly, we propose a 2-level PCAT hierarchy, an L1 PCAT (PCAT1) coupled to an L1 instruction cache (L1i)
 and an L2 PCAT (PCAT2) coupled to an L2 cache.
Like an L1i, the PCAT1 is \textit{virtually-indexed physically tagged (VIPT)}, though it could also be \textit{physically-indexed physically tagged (PIPT)} as it's output isn't needed until later. Moreover, the decision to access the PCAT1 is governed by a \textit{access PCAT} bit, which is obtained alongside L1i access, as we see below. Like an L2 cache the PCAT2 is PIPT.

Entries are created in the PCAT1 when a load instruction that misses in the CDTLB retires.
The PC of the load instruction (or the cache block address of the cache block holding the instruction in an alternate design) is used to determine a place (set and way)
where the new entry will reside, and the tag bits populated.
The PTE returned on the CDTLB miss, and the VPN of the accessed data address,
are entered into the VPN-PTE bits of the entry, and the Useful bit is unset. 

When an entry in the PCAT1 provides a useful translation,
the Useful bit gets set.



A newly-created PCAT1 entry moves to PCAT2 when it is evicted from PCAT1 thus creating a new entry in the PCAT2, eventually populating the PCAT2. An alternative is to create an entry in the PCAT2 at the same time as in PCAT1.  Eviction from the PCAT1 is independent of eviction of the corresponding code cache block from the L1i. 
However, an entry is moved from PCAT2 to PCAT1 (resulting in an eviction of an entry from PCAT1) when the corresponding instruction block is moved from the L2 to the L1i, and not on a PCAT1 miss, as in a typical cache hierarchy. 


\begin{figure}[t]
  \centering
  \includegraphics[width=1.0\linewidth]{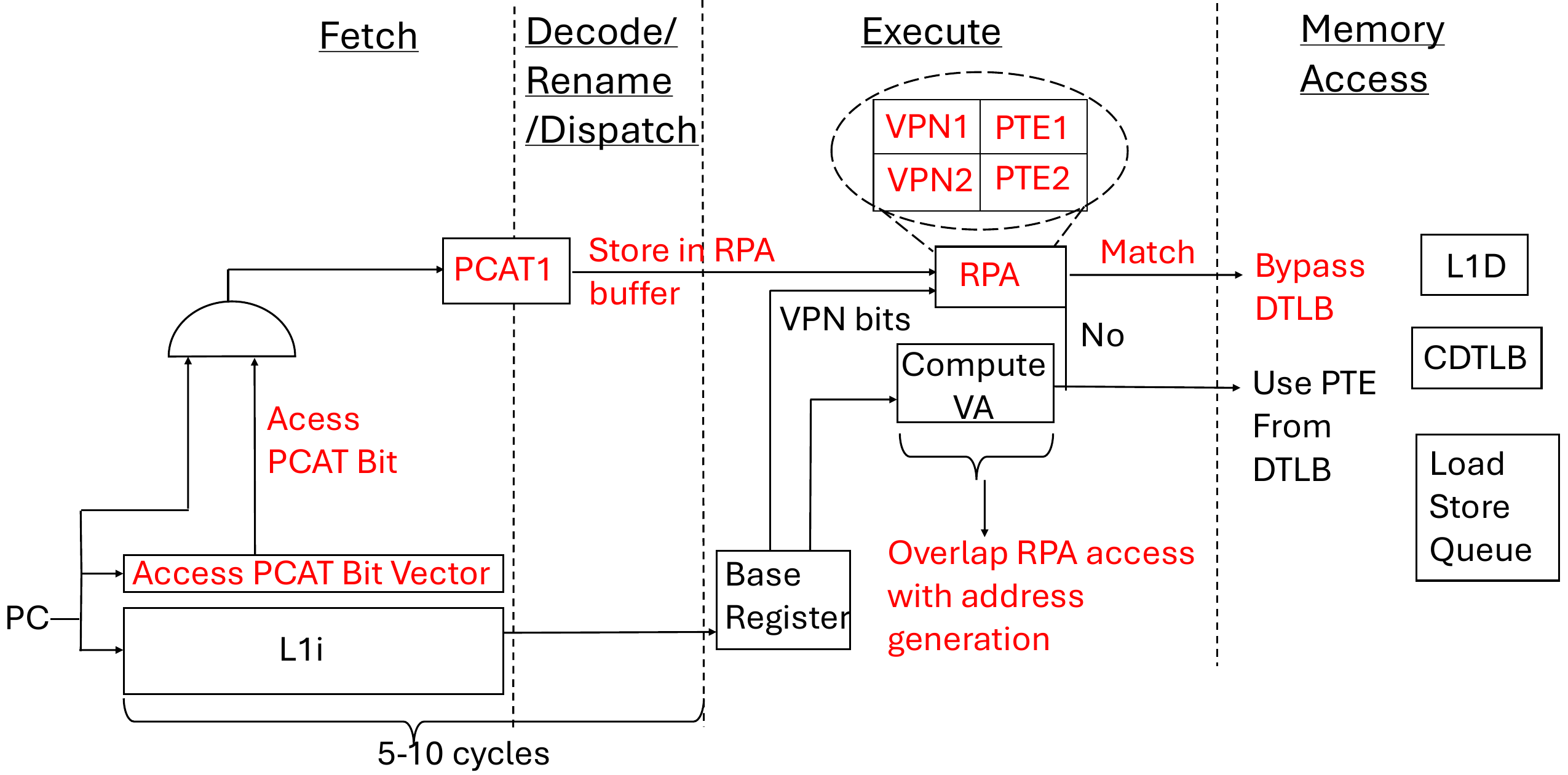}
	\caption {Relevant Portion of Load Operation}
  \label{fig:load-insn-pipeline}

\end{figure}



\subsection {Load Pipeline with PCATs}
\label{sec:pcat1-pcat2-access}

Figure \ref{fig:load-insn-pipeline} illustrates the relevant stages involved in the operation of a load instruction within the pipeline. PCAX-specific operations are highlighted in red for clarity.
The PCAT1 is accessed using the PC of a load instruction, alongside the fetch of the load instruction.
However, we only want to access the PCAT1 for loads for which there is an entry in the PCAT1,
and not for every load instruction. The latter is wasteful as most load PCs
will not have a matching entry in the PCAT. 
To avoid this, we maintain a (pre-decoded) \textit{access PCAT} bit with each instruction in the L1i to indicate whether the instruction PC was that of a problem load instruction and was used to create a PCAT entry. Since only a small subset of instructions typically fall into this category, these bits are rarely set.
Indeed, even for cache blocks that contain a problem load, often there is only one problem load and we can use an access PCAT bit per block, rather than per instruction, in the L1i.
The access PCAT bit is set for the corresponding cache block when the PCAT1 entry is created, which is when the load that misses in the CDTLB retires (as described in the Section \ref{sec:pcat1-struc-op}). 
The PCAT1 is accessed with the instruction PC (or cache block address of the cache block holding the instruction), when the corresponding access PCAT bit set.

 
Though the PCAT1 is accessed alongside instruction fetch, its output---a VPN-PTE---is not needed until much later
in the processing pipeline.  In particular, until the address for the data of the load
instruction has been determined (and the VPN known).
To reduce access energy (at the expense of latency), we adopt the well-known two-step process for a cache access. The tag array is accessed first to check for a hit. If a hit occurs, the data array is accessed in the second step.
On an access and a match, the PCAT1 provides a VPN-PTE, which is taken to the
data address generation stage of the pipeline and maintained in a \textit{Recent PCAT Accesses Buffer (RPA buffer)}.

The RPA buffer is 1-2 entries, fully associative and is accessed using the data VPN. It acts like a nano-TLB, but with important differences: (i) it is only populated by select load instructions, those that hit in the PCAT1, and only with a VPN-PTE entry from the PCAT1, unlike a traditional nano-TLB which can be populated by any load instruction from a regular TLB, and (ii) it uses a FIFO replacement if there are multiple entries for a new entry from (i), and not a replacement scheme that is influenced by accesses.  Similar to a nano-TLB, it can be accessed by any load instruction, using the VPN of the data address of the load instruction.
The RPA buffer supports the same access rate as the CDTLB.

The primary intent of the RPA buffer is to hold the VPN-PTE obtained from the PCAT1 by a (problem) load instruction, until the time the data address has been calculated for that (problem) load instruction, to see if the VPN-PTE is the correct one for the (problem) load instruction (the VPNs match), and whether or not the CDTLB needs to be accessed.  However, an entry in the RPA buffer, while intended for use by the load instruction whose PC was used to access the PCAT1, can also be used by other instructions that were not involved in a PCAT1 access.
As others have observed,
different load instructions that are close in the instruction stream, and that use the same base register, often access the same virtual page (and PTE) \cite{austin1996high,chiueh1992eliminating}. 
Similarly we have observed that different static loads, likely accessing different fields of a data structure (same base register and different offsets), are temporally close dynamically, and access the same PTE.  However, they are not all contiguous in the dynamic stream; they are interleaved with other instructions accessing other VPNs (that hit in the CDTLB).
We have found this RPA buffer, which fleetingly holds \textit{select PTEs}, to service the address translations for a fair number of instructions, as we shall see in section \ref{sec:ServiceSites}.

Next we consider options for how and when the RPA buffer is accessed.  An obvious option is to access it once the data address for a load instruction has been calculated, with the VPN bits from the calculated address.  Noting, however, that the VPN-PTE obtained from the RPA buffer is just an optimization, with a fall back to the CDTLB, the buffer could also be accessed with a set of VPN bits that are \textit{highly likely} to be the VPN bits of the data address.  The VPN bits needed to determine a match could be some other set of bits involved in the data address calculation. 
Accordingly, when the base register for a load instruction is read for a data address calculation (or even earlier depending upon the pipeline) the VPN bits of the base register are compared with the VPN bits of the RPA buffer entries.  
On a match, the PTE from the RPA buffer is used if the VPN does not change as a result of the address calculation (i.e., offset smaller than a page and no carry into VPN bits).  In this way the access of the RPA buffer can be carried out in parallel with (or prior to) the address calculation. If the address calculation results in a change in the VPN (an event that is expected to be quite infrequent), the entry is not used.
The obtained VPN-PTE also contains the permission bits for the page (like a CDTLB) and permission checks are made at this point. 
If there is no match (or the VPN bits change as a result of address calculation) the data address is submitted to the CDTLB to obtain the corresponding PTE.

The RPA buffer has as many ports as the CDTLB. It is invalidated whenever the PCAT is invalidated; it need not be invalidated for other events that require a pipeline flush.



\subsection {Address bits for PCAT access}



The bits used for a PCAT access could be from the PC of a (problem) load instruction, or the bits of the cache block number containing a (problem) load instruction.  The former allows more flexibility as an arbitrary number of loads from the same cache block can reside in a PCAT.  However, this requires a per-instruction access PCAT bits in the L1i cache and also multiple accesses to the PCAT1 if these loads reside in different sets of the PCAT1.  In the latter case, there needs to be only one access PCAT bit, and one PCAT1 access per cache block.  Multiple VPN-PTE entries for multiple problem loads in a cache block can reside in the different ways of a set and thus be obtained in parallel, with the set associativity capping the number of problem loads per cache block for which the PCAT1 can maintain information. 
The usage of bits of cache block number also simplifies the access of PCAT2 tables and the movement of entries from the PCAT2 to PCAT1, as the same cache block number can be used to access both the L2 cache and the PCAT2, and the entries transferred when the access PCAT bit of the cache block in L2 is set. We briefly evaluate the two options quantitatively in section \ref{sec:PCATIndexingEval}. 



\subsection {PCAT/CDTLB/STLB and Page Table Interactions}


Like a  CDTLB, a PCAT operates like a cache of PTEs, albeit one that
is accessed with a load PC rather than a data address. When PTEs are cached, appropriate actions need to be taken to ensure correctness.
When an update is made in a PTE, copies cached in CDTLBs must be invalidated using a \textit{TLB Shootdown} \cite{teller1987tlb,pham2017tlb}.
These changes include the PPN mapping, permissions, and potential updates to the Access bit.


Since the PCAT is accessed with a load PC, and not with a data virtual address, operations using a data virtual address get more involved. Instead of invalidating individual entries using data virtual addresses (as in TLB shootdown)
we invalidate the entire PCATs on any change to an
entry in the page table (a \textit{PCAT Shootdown}) as well as the RPA buffer.
Since a PCAT entry is created on a (first) CDTLB miss, where the page walker also updates the Access bit in the page table and gives the correct value
to the CDTLB, the PCAT will also have the correct value of the Access bit.  
Any subsequent clearing of the Access bit in the page table (by the OS) results in a PCAT Shootdown.
Thus having a PCAT does not impact the operation of the OS page replacement algorithm.

Handling the Dirty bit requires special consideration because the first store to a page may occur after the corresponding entry has already been inserted into a PCAT. Since stores do not consult or update PCAT structures, the Dirty bit recorded in a PCAT entry can become stale even though the Dirty bit in the architectural Page Table and TLB hierarchy has been set. Moreover, a given PTE may appear in the PCATs with different load PCs, further complicating coherence.

A straightforward solution would be to eagerly invalidate the PCATs on any PTE modification. Although correct, this approach would introduce unnecessary invalidations and reduce the performance benefits of PCAX. Instead, we tolerate temporal incoherence in PCAT Dirty bits and rely on the existing TLB hierarchy to ensure correctness.

In a conventional hierarchy, the STLB is checked before invoking a page walk. PCAX extends this mechanism to safely absorb Dirty‑bit incoherence originating from PCATs. Consider the case where the PTE inserted into the CDTLB from PCAT1 has the Dirty bit unset because the PCAT entry was created prior to the first store. That first store has already updated the Dirty bit in the Page Table and propagated it to the STLB and CDTLB (under the standard non-inclusive, non-exclusive rules). If a later load brings an incoherent version of the PTE into the CDTLB from PCAT1, a subsequent store will set the Dirty bit in the CDTLB and attempt to propagate the update toward the STLB. The propagation stops early if the STLB already contains the PTE with the correct Dirty bit set. Thus, correctness is preserved without requiring a page walk. A page walk is only required if the STLB either misses or also contains an unset Dirty bit.
At the next periodic PCAT invalidation (Section \ref{sec:PeriodicInvalidation}), the PCAT entries are refreshed with up‑to‑date PTE state, and the above events no longer occur.
In our experimental evaluation, with a periodic invalidation interval of 1M instruction,
for a small number of benchmarks there are few tens of additional STLB accesses, and a small number 
($<$ 5) additional page walks.
Thus, PCAX preserves correctness while minimizing PCAT invalidations through controlled tolerance of temporal Dirty-bit incoherence.

\subsection {PCAX Coupling with the CDTLB/STLB}

An issue is whether PCAT1/PCAT2 (for PCAX) and CDTLB/STLB (for conventional address translation) should be coupled.
That is, when a PTE is brought into the PCAT1, as a result of a CDTLB/STLB miss,
should the PTE also be placed in the CDTLB/STLB (the structures are coupled), 
or solely in the PCAT1 (not coupled).
Not coupling frees up capacity in the CDTLB/STLB which could be used for other
translations whereas not placing a PTE in the CDTLB when it is
brought into the PCAT1 results in misses for other proximate load instructions accessing the same PTE. 
Such instructions cannot get the translation from the PCAT1
(because of different PCs) and it will not be there in the CDTLB.

Our experimental observations 
suggest that PCAT1/CDTLB coupling is beneficial for most applications. 
Accordingly, when a PTE is placed in the PCAT1, it is also placed in the CDTLB.
For the STLB, given that there is little reason to maintain inclusion between
the STLB and the CDTLB/PCATs, we do not couple the PCATs and the STLB.
Empirically we have observed that this reduces duplication in the PCAT1/CDTLB and STLB, 
leading to a slightly lower STLB miss rate.


\subsection {Replacement in the PCATs}
\label{sec:repl-pcat2}

The PCATs can use normal cache replacement policies to determine a victim entry.  However, we have found it beneficial to enhance a normal (e.g., LRU) replacement policy
with some additional information.
To preserve entries that have been deemed to be useful
over those that have not been deemed so, we use the Useful bit to exclude
an entry from a replacement when possible.
Likewise, we have found it beneficial to try to preserve entries that were created as
a result of an STLB miss, especially for the PCAT2, since they can potentially provide a translation that might
otherwise have to be obtained via an STLB miss. 

The STLB Miss (STLBM) bit indicates such entries. The modified LRU replacement algorithm that we employ does not choose an entry with either its Useful or STLBM bit set, unless it doesn't have a choice. 
We now describe how the STLBM bit is set. Initially on cold start every entry in the PCATs (and CDTLBs and STLBs) comes from an STLB miss so every entry has its STLBM bit set. However, the PCATs are periodically invalidated, as we discuss in the next section.  Invalidation of the PCATs does not involve or impact the CDTLB/STLB, whose contents remain as they are. After an invalidation, when a PCAT entry is created, the STLBM bit is set on not based upon whether the entry came from an STLB miss or hit, with some (not all) entries having their STLBM bit set, and allowing us to use it in the replacement policy.


\subsection {Periodically Invalidating the PCATs}
\label{sec:PeriodicInvalidation}

We periodically invalidate all PCAT structures to prevent entries from persisting indefinitely and causing unproductive behavior. Earlier, we described two scenarios that benefit from periodic invalidation: (1) cold-start behavior affecting the STLBM bit, and (2) temporally incoherent Dirty-bit interactions that can lead to unnecessary page walks. We have also identified a third, subtler pathology that arises from code blocks that become cold.

Consider an instruction cache block A containing code for multiple control-flow paths. Suppose a problem load is detected along one such path, causing a PCAT entry to be created and its Useful bit to be set. If that execution path later becomes cold, the corresponding PCAT2 entry continues to occupy a way in its set. Meanwhile, other problem loads from unrelated code blocks map to the same PCAT2 set and occupy the remaining ways. When block A is reinserted into the L1 instruction cache, its PCAT2 entry is migrated to PCAT1. Importantly, this migration causes the PCAT2 entry to be marked as “accessed” by the local replacement policy—even if the entry is no longer used by the processor while resident in PCAT1. If A is brought into the L1i more frequently than other code blocks that map to the same PCAT2 set, and the PCAT2 replacement policy relies solely on local access information, then the cold entry from A is effectively shielded from eviction. As a result, useful entries belonging to other code blocks are preferentially evicted, while the cold PCAT2 entry persists far longer than desirable. One potential solution is to incorporate actual use information gathered while the entry resides in PCAT1 into the PCAT2 replacement algorithm. However, this approach introduces additional complexity and complicates the replacement logic. Instead, PCAX adopts a simpler and robust strategy: periodically clearing the PCATs. Periodic invalidation flushes stale or cold PCAT entries, allowing the structures to be naturally re-populated with entries corresponding to the currently active working set. This approach effectively eliminates long‑term occupation by cold entries and prevents replacement imbalance. Similar periodic clearing strategies have been shown effective in other microarchitectural structures such as memory dependence predictors—where learned store–load relationships that have gone cold are periodically discarded and relearned \cite{chrysos1998memory,moshovos1997dynamic}.

\subsection {Selectiveness of Loads Using PCAT}

PCATs are intended to assist CDTLBs by providing translations that might miss in the CDTLB.
PCAT sizes can be kept small because the number of static loads that miss in a CDTLB are small.
However, PCAX could also be used for a small number of other "high value" loads from the set of
loads that would normally hit in the CDTLB (whose number is, of course, much larger).
Like the data of figure \ref{fig:prev-page-accesses-misses-server-64-256-1k-tlb} for loads that miss in a CDTLB,
we have also found that frequently a dynamic instance of a load that hits in a CDTLB
uses the same PTE as the last dynamic instance
(data in section \ref{sec:BenchChars}).
Judiciously using PCAX for loads that might normally hit in a CDTLB, with reasonably small-sized PCATs,
to improve translation latency and energy is left for future work.




\section{Evaluation}
\label{sec:Eval}

\begin{table}[]
\centering
\caption{Simulated Machine Parameters}
\scalebox{0.9}{
\begin{tabular}{ll}
\hline
	\multicolumn{2}{c}{\textbf{Processor}}                \\ \hline
        Fetch Width & 6 instructions \\
Branch Predictor     & Hashed perceptron                       
  \\
Execute width        & 4 instructions                          \\
Retire width         & 5 instructions                          \\
Re-order buffer      & 352 entries                              \\
	Load, store queue    &  128, 72 entries                         \\ \hline

\multicolumn{2}{c}{\textbf{Memory   hierarchy}}                                                                                                          \\\hline
CPU Frequency                                                                          & 4GHz                     \\\cline{2-2}
L1 ITLB                                                                            & 64-entry, 4-way, 1-cycle, 4-entry MSHR                     \\\cline{2-2}

\multirow{3}{*}{L1 DTLB} 
  & 64-entry, 4-way, 1-cycle, 4-entry MSHR, \\ 
  & 2 lookups per cycle, 2 AGU \\ \cline{2-2}
\multirow{2}{*}{L2 TLB}                                                            & 1536-entry, 12-way, 8-cycle;                               \\
                                                                                   & 4-entry MSHR, 1 page walk / cycle                          \\\cline{2-2}
\multirow{3}{*}{\begin{tabular}[c]{@{}l@{}}Page  Structure \\ Caches\end{tabular}} & 3-level Split PSC, 2-cycle;  \\ 
                & PML4: 2-entry, fully,   PDP: 4-entry, fully; \\ &
                    PD: 32-entry, 4-way                  \\\cline{2-2}
\multirow{2}{*}{L1 Dcache}                                                            & 48KB, 12-way, 5-cycle, 8-entry MSHR;                               \\
                                                                                   &  next line prefetcher                         \\\cline{2-2} 
\multirow{2}{*}{L2 Cache}                                                            & 512KB, 8-way, 10-cycle, 16-entry MSHR;                               \\
                                                                                   &  spp-dev prefetcher                        \\\cline{2-2}
LLC                                                                                & 2MB, 16-way, 20-cycle, 32-entry MSHR                       \\\cline{2-2}
\multirow{2}{*}{DRAM}                                                            & 4GB, one 8-byte channel,  1600 MT/s;                               \\
                                                                                   &  tRP=tRCD=tCAS=12.5                          \\\cline{2-2} 
\end{tabular}
}
\label{fig:sim-params}
\end{table}

\subsection{Benchmarks and Experimental Infrastructure}
\label{sec:Infra}

For our evaluation, we use ChampSim \cite{gober2022championship}, a detailed simulator that simulates an out-of-order core with 6-wide fetch and 4-wide issue.   
Table \ref{fig:sim-params} summarizes the parameters of the simulated microarchitecture.
This version of ChampSim also models a realistic page table walker as found in x86 architectures. Key aspects modeled include: (i) variable latency cost of page table walks and (ii) caching of page walk references in data memory hierarchy as in \cite{vavouliotis2021exploiting}. Specifically, this models a 4-level page table, a hardware page table walker, and a 3-level split page structured cache. The page walker supports up to 4 concurrent STLB misses.  All caches and TLBs use a normal LRU replacement policy.  The default page size is 4KB and the \textit{default PCAX scheme} employs a 64-entry, 2-way PCAT1, a 2K entry, 8-way PCAT2, and a 2-entry RPA Buffer. 
PCAT1 uses normal LRU while PCAT2 uses the slightly modified LRU replacement which was explained in Section \ref{sec:repl-pcat2}.

For benchmark programs, we use a set of 84 server traces from Qualcomm, which were released after modifications recently \cite{feliu2023rebasing,feliu2023rebasing1}.
 These traces have been widely used in recent papers using the Champsim simulation infrastructure, and studying a variety of issues in instruction supply, memory hierarchies, and microarchitecture \cite{vavouliotis2021exploiting,vavouliotis2021morrigan,asheim2023storage,ros2021cost,kim2017kill} and more which can be found in \cite{gober2022championship}. 

These traces have a data footprint of 20MB–100MB, which is smaller than those used in other address translation studies \cite{karakostas2016energy,karakostas2015redundant,gosakan2023mosaic}. However, the data access patterns produce CDTLB miss rates from 1\% to 31\%, and CDTLB MPKIs from 1 to 53, with a 64-entry, 4-way associative data TLB. This makes them well suited for evaluating PCAX. Many benchmarks traces have 100M instructions, while some of the traces are smaller (30M instructions) \cite{traces}. These traces, released by Qualcomm, have been used in recent works focused on L1 cache and TLB behavior \cite{ros2021cost,vavouliotis2021exploiting,vavouliotis2021morrigan,chasapis2025instruction}.


\subsection{Storage Costs}




A PCAT entry requires approximately 11.6 bytes: a 20-bit tag, 1 valid bit, 1 useful bit, 1 STLB Miss bit, a 38-bit VPN, and a 4-byte PTE. A 64-entry PCAT1 occupies roughly 744B. PCAT2 configurations with 512, 1024, and 2048 entries require approximately 6KB, 12KB, and 24KB, respectively.

Access PCAT bits are stored per instruction cache block: a 32KB L1 instruction cache with 64-byte blocks requires 512 bits (64B), while a 512KB L2 cache requires 8192 bits (1024B). It would take 2048B with a 1MB L2 cache. The RPA buffer adds an additional 18 bytes.

In our baseline configuration, we use a 64-entry PCAT1 and a 2048-entry PCAT2, resulting in a total storage cost of approximately 25.8 KB. However, as demonstrated in Section \ref{sec:CDTLBReduction}, smaller PCAT2 configurations with 512 and 1024 entries—requiring 7.8 KB and 15.8 KB respectively—achieve similar performance across a range of applications.

\subsection{Results Overview}
We start our evaluation by presenting some relevant characteristics of the benchmark programs in section \ref{sec:BenchChars}. Then section \ref{sec:CDTLBReduction} evaluates the reduction in MPKIs for primary DTLBs for different PCAT configurations. Section \ref{sec:large-page-sizes} studies the effect on larger page sizes. 
Section \ref{sec:ServiceSites} presents the
service sites for data address translation for a select benchmark set and then sections \ref{sec:CPIBenefit} and \ref{sec:EnergyImpact} evaluate the impact on CPI and energy, respectively. Lastly, Section \ref{sec:FlushPCAT} studies the effect of invalidating the PCATs periodically.

The ordering of the benchmarks on the X-axis in the different figures is different for better visual presentation. Most of our evaluations present results for individual benchmarks.  Average results are presented using the \textit{geometric mean (GM)}.
%

\subsection{Benchmark Characteristics}
\label{sec:BenchChars}



\begin{figure}[h]
 \includegraphics[width=0.9\linewidth]{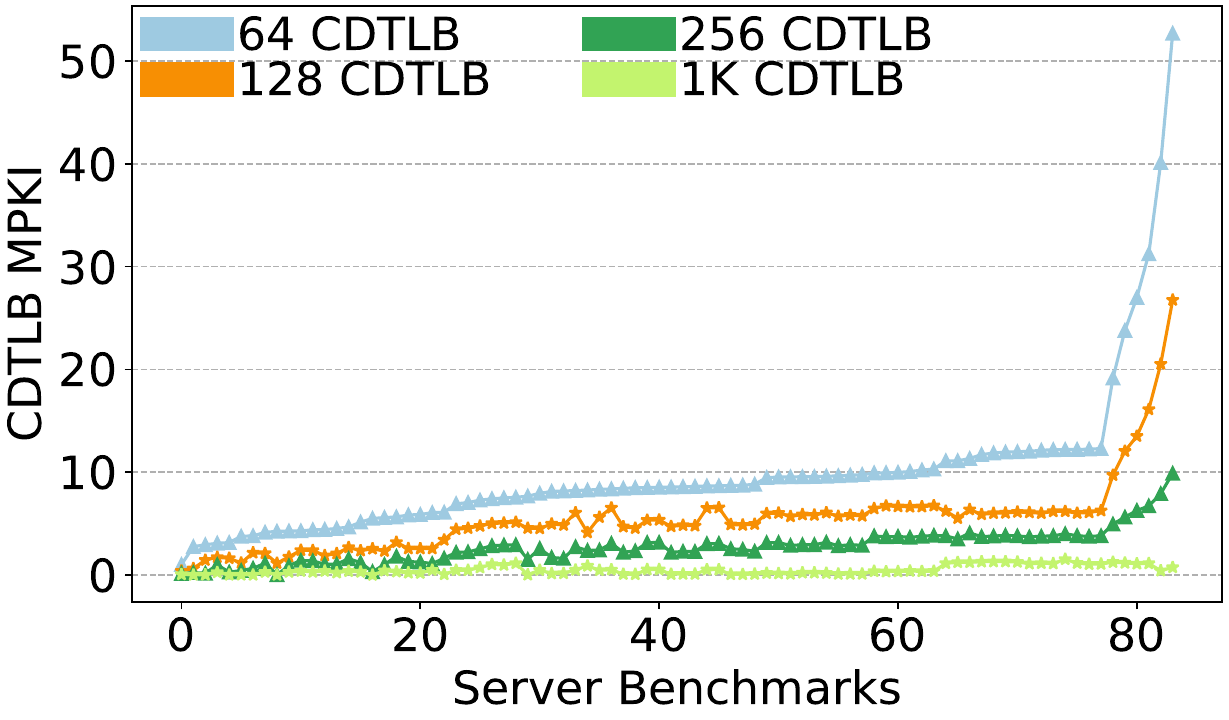}
	\caption {CDTLB Load MPKI}
  \label{fig:tlb-mpki-varying}
\end{figure}

Figure \ref{fig:tlb-mpki-varying} shows the CDTLB Load MPKI for CDTLB sizes of 64, 128, 256, and 1024 entries, with 4-way associativity in each case.
For a 64-entry CDTLB, most of the applications have a MPKI $>$ 7.
Even for a (single-level) 1024-entry CDTLB, many applications have a MPKI $>$ 1.


\begin{figure}[t]
  \includegraphics[width=0.9\linewidth]{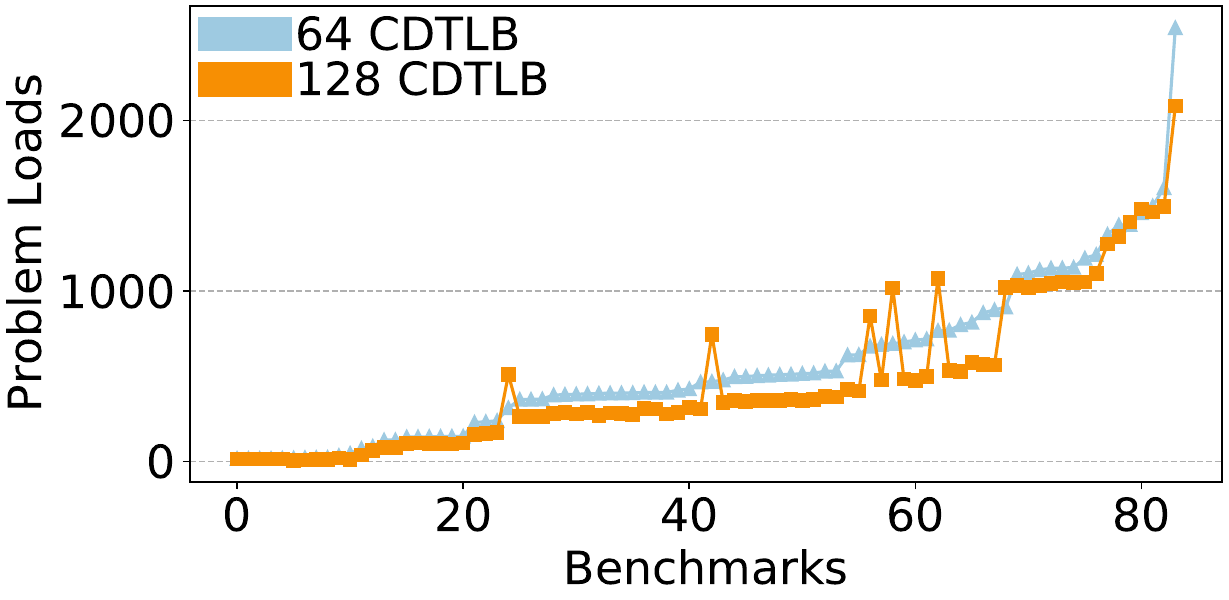}
	\caption {Number of Problem Loads}
  \label{fig:per-static-load-server}
\end{figure}

\begin{figure}[t]
  \includegraphics[width=0.9\linewidth]{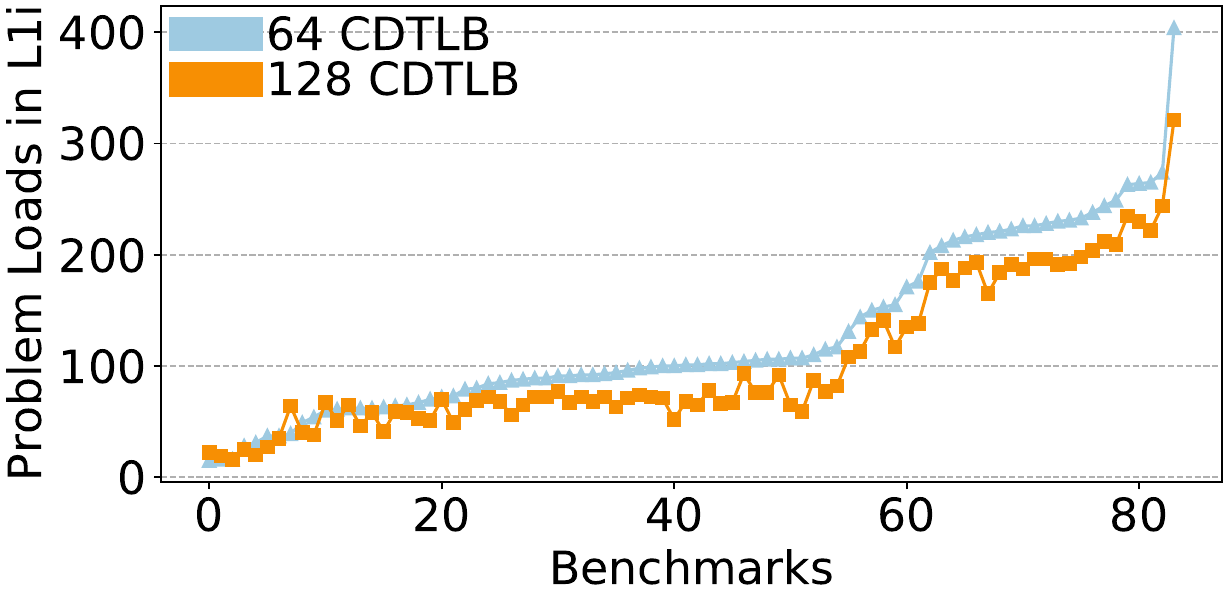}
	\caption {Maximum Problem Loads in L1i}
  \label{fig:blks-problem-pc-icache-qual}
  
\end{figure}

\subsection{PCAT Table Sizes/Hierarchy}

Figure \ref{fig:per-static-load-server} shows the total number of problem loads for 64 and 128 entry CDTLBs.  The total number is of the order of a few hundred and less than 2000 in almost all cases. 
These numbers give an idea of how large of a PCAT2 might work well.
In some cases, the number of problem loads is higher for the 128-entry CDTLB than for the 64-entry configuration. This occurs in applications with a high MPKI with a 64-entry CDTLB (e.g., 20), which drops significantly (e.g., to 10) with the larger CDTLB. Accounting for 90\% of the misses encompasses more problem PCs,
which are not included in the 90\% when the miss rate is higher.


Since the PCAT1 needs to accommodate only problem loads that are resident in the L1i, we consider this number in Figure \ref{fig:blks-problem-pc-icache-qual}.  The figure presents the \textit{maximum} number of problem loads resident in the 32KB L1i at any time for 64-entry and 128-entry CDTLBs.  This maximum number is in the small hundreds.
The average number (not shown) is smaller.
As we shall see, PCAT1 sizes that can accommodate a few tens of problem loads
(e.g., 64) are adequate in most cases.

Taken together, these observations motivate a hierarchical PCAT organization. Although the cumulative number of problem loads across an application can reach into the thousands—suggesting the need for a reasonably sized PCAT2—only a small subset of these loads is resident in the L1i at any given time. This allows a small, fast PCAT1 to track the active working set of problem loads in the L1i, while a larger PCAT2 retains translations for loads that are evicted from the L1i but likely to be fetched again on an L1i miss.

\subsection{PCAT Indexing}
\label{sec:PCATIndexingEval}

\begin{figure}[t]
 \includegraphics[width=0.9\linewidth]{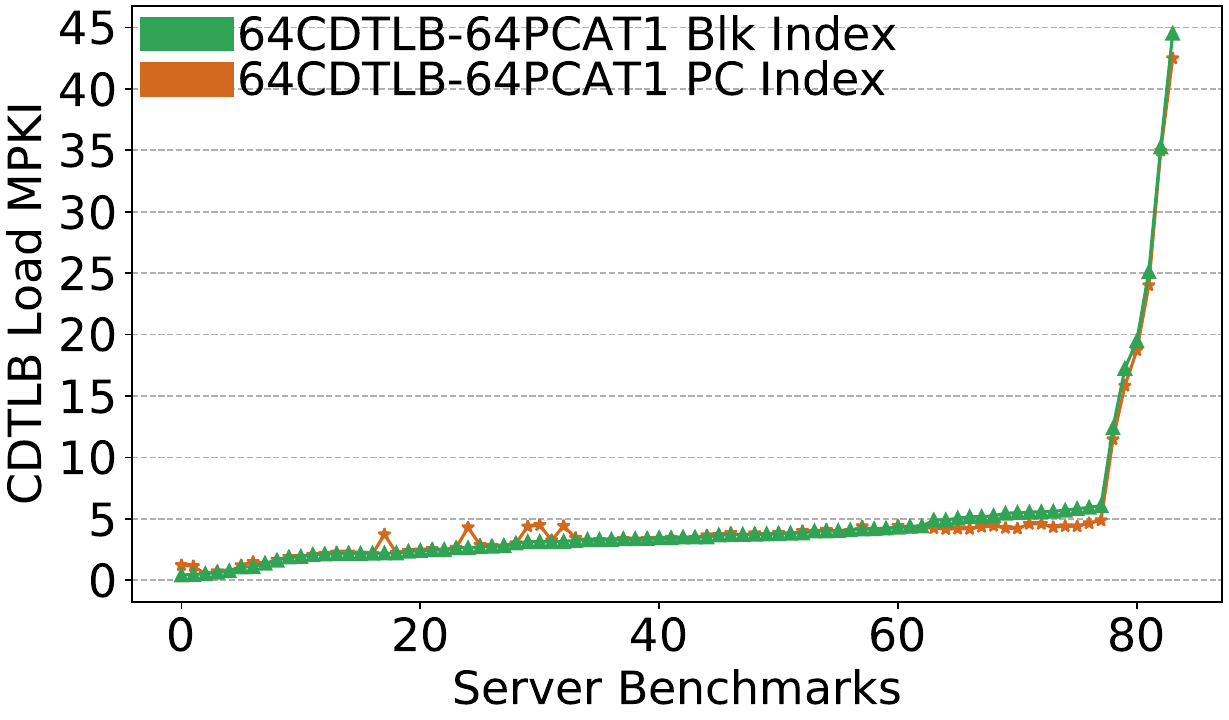}
	\caption {PCAT Indexing}
  \label{fig:pcat-indexing}
\end{figure}

Figure \ref{fig:pcat-indexing} presents the CDTLB MPKIs for cache block address and load PC indexing of a PCAT. The MPKIs are very similar in most cases, with either performing slightly better in some cases.  We use the block address indexing for all the results we present subsequently because it results in only a single PCAT access for a block of instructions.

\subsection{Reduction in Primary CDTLB Misses}
\label{sec:CDTLBReduction}

\begin{figure}[t]
  \includegraphics[width=0.9\linewidth]{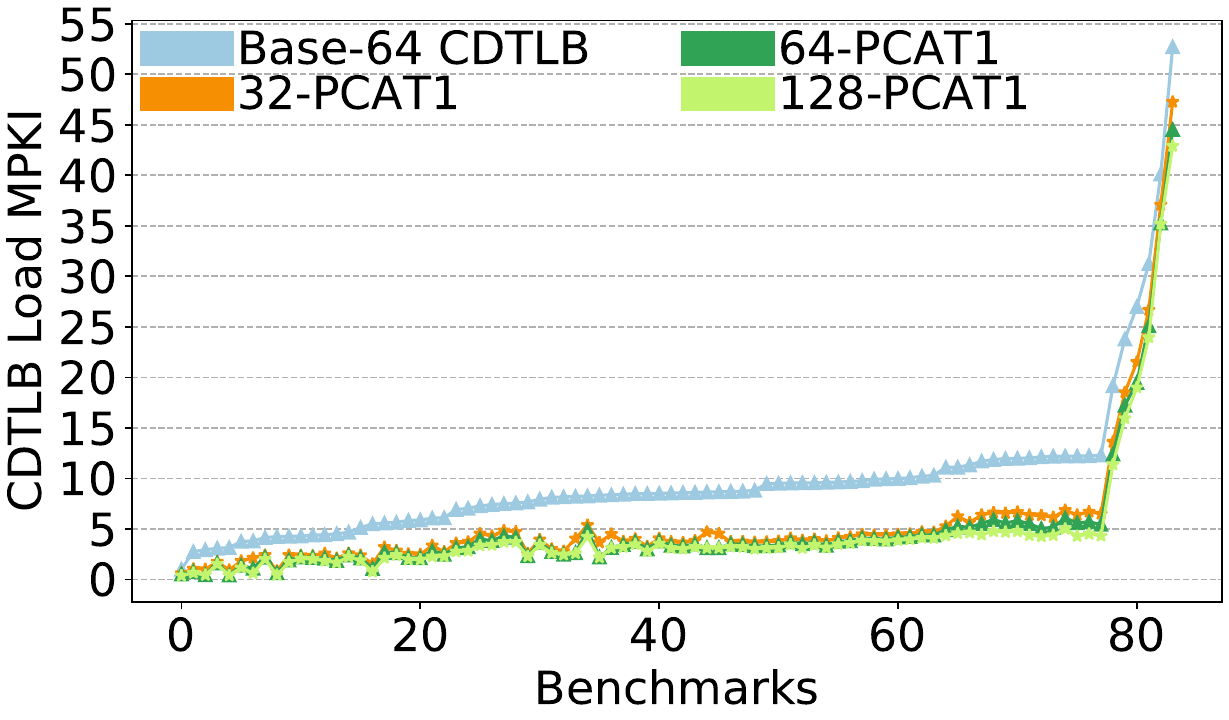}
	\caption {Effective Load MPKI (64-Entry CDTLB)}
  \label{fig:per-64tlb-qual-server}
\end{figure}

\begin{figure}[t]
  \includegraphics[width=0.9\linewidth]{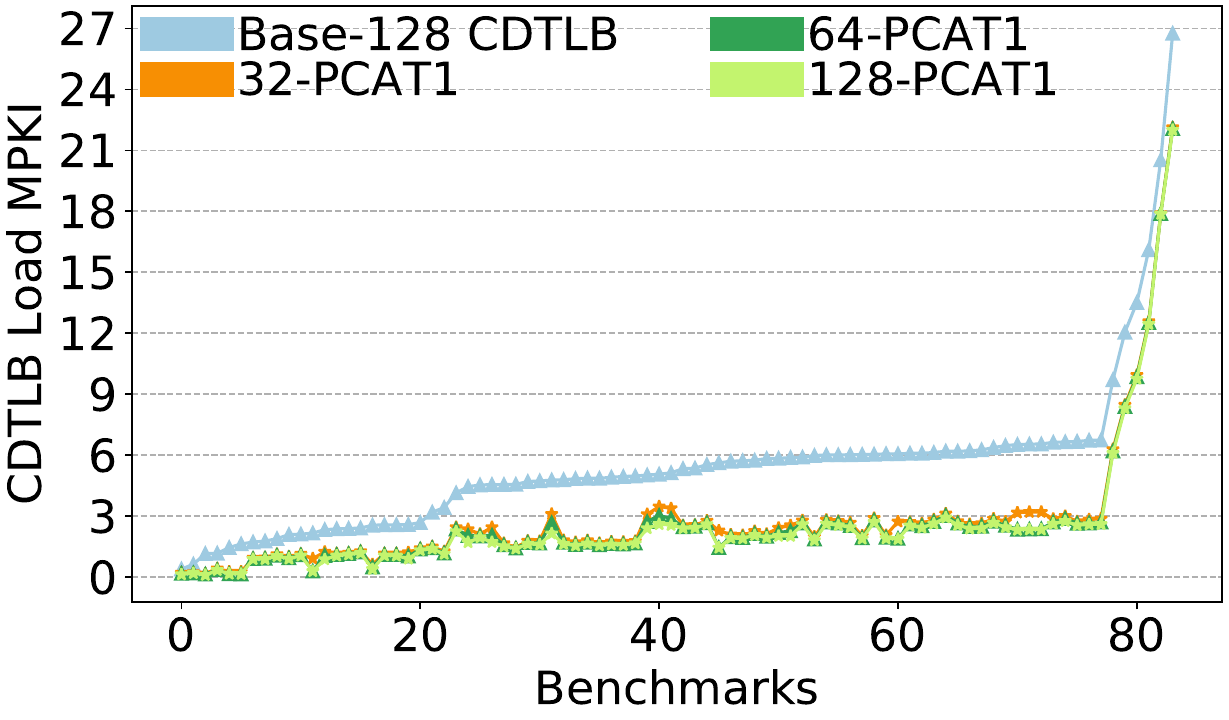}
	\caption {Effective Load MPKI (128-Entry CDTLB)}
  \label{fig:per-128tlb-qual-server}
\end{figure}

We now consider the reduction in the primary CDTLB misses due to the translation requests fulfilled by a PCAT1. Figures \ref{fig:per-64tlb-qual-server} and \ref{fig:per-128tlb-qual-server} show the effective CDTLB MPKI for 64- and 128-entry, 4-way set associative CDTLBs, respectively.  The effective MPKI is the MPKI for load references that can't be serviced by the PCAT1 and CDTLB, and end up accessing the STLB.
In each case, we plot the base CDTLB MPKI (Base) and the effective MPKI with 32-, 64-, 128-entry, 2-way set associative PCAT1s.  The PCAT2 size is 2048 entries and is 8-way set associative in all cases.
Detailed data for each benchmark is shown in the figures for the above PCAT1 configurations but below we also report geometric mean (GM) averages for 8-way and fully-associative configurations for which detailed data is not presented in the figures.


\begin{figure}[t]
  \includegraphics[width=0.9\linewidth]{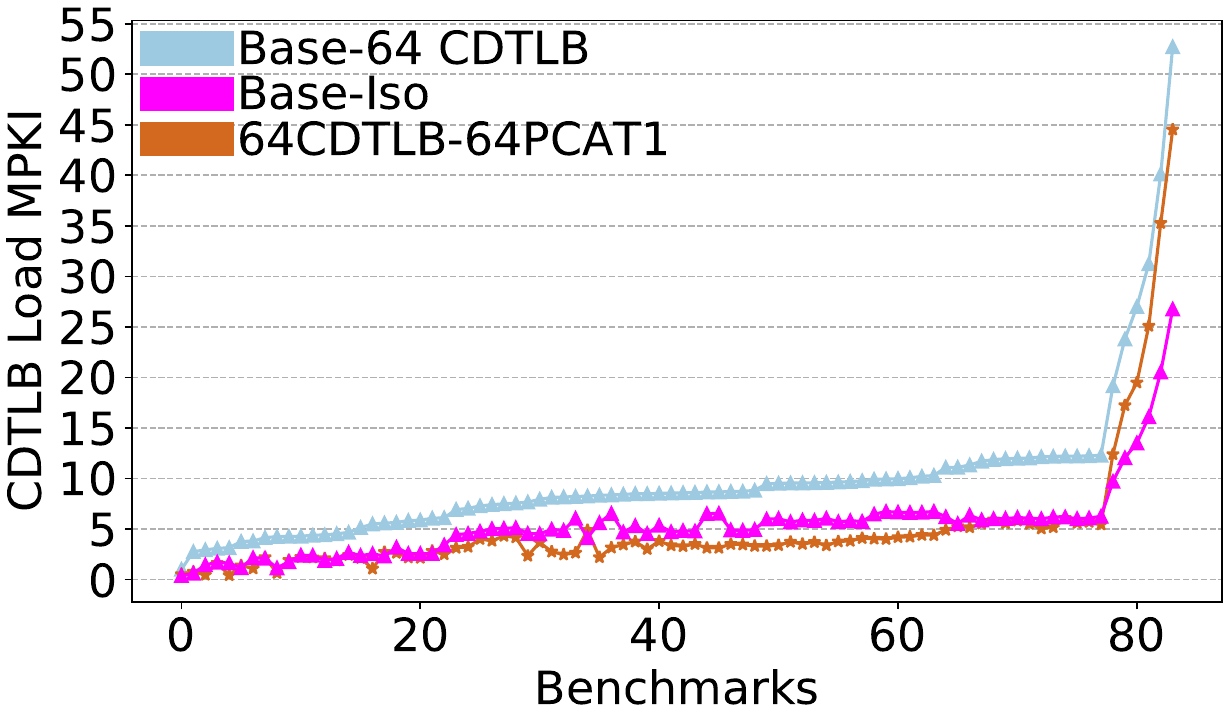}
	\caption {Comparing Base-Iso with PCAX}
 \label{fig:iso-storage-study}
\end{figure}


\begin{figure}[t]
  \includegraphics[width=0.9\linewidth]{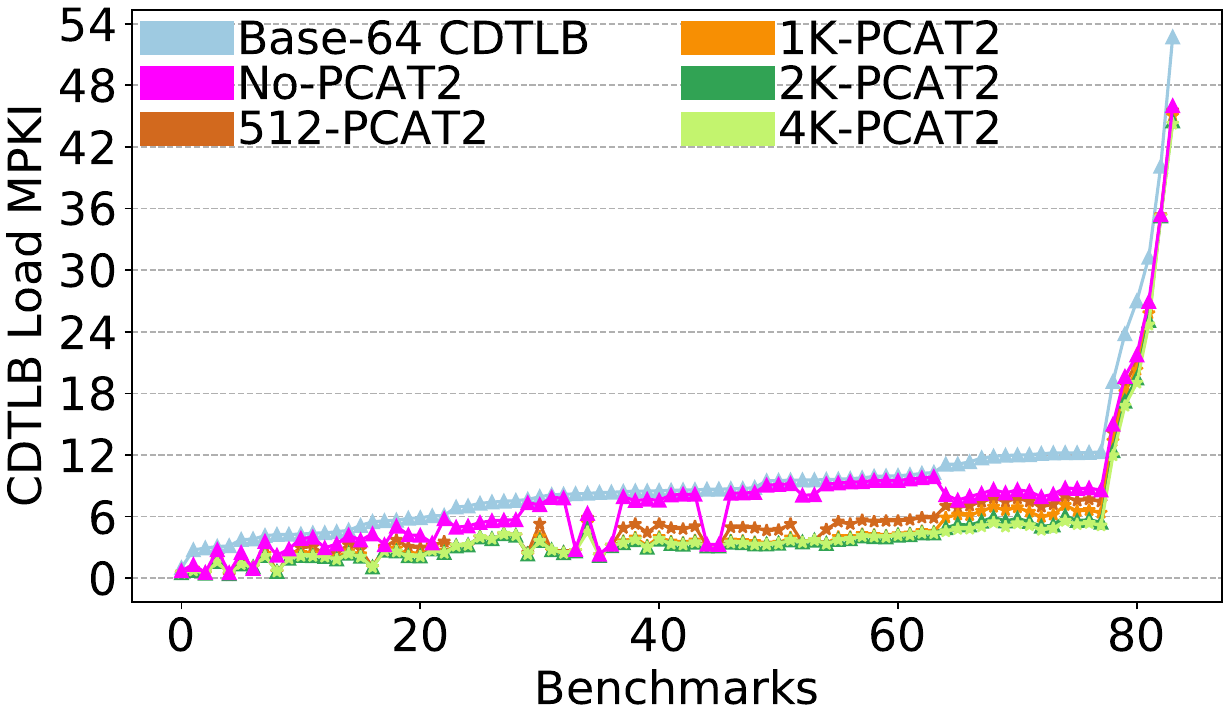}
	\caption {Varying PCAT2 size}
  \label{fig:per-64tlb-qual-server-pcat2}
\end{figure}


The figures demonstrate a substantial reduction in the effective CDTLB MPKI across nearly all workloads, even with a modest 32-entry PCAT1. Increasing the PCAT1 to 64 entries provides additional reductions for several applications, although benefits beyond 64 entries are marginal. The magnitude of improvement is noteworthy. From Figure \ref{fig:per-64tlb-qual-server}, observe that while the baseline MPKI ranges from 5-13 (geometric mean (GM): 8.19) for most applications, a 64-entry PCAT1 lowers this to 2-5 (GM: 3.43). For 10 benchmark traces, the effective MPKI is reduced by approximately 3X, and for 68 traces it is reduced by more than 2X. Similar improvements occur with an 8-way associative CDTLB: the baseline CDTLB MPKI GM of 6.75 is reduced to 2.64 with PCAX. With a fully associative CDTLB, the CDTLB MPKI GM decreases from 4.85 (baseline) to 2.23 with PCAX. Significant gains persist even with a 128-entry CDTLB, as shown in Figure \ref{fig:per-128tlb-qual-server}. For 67 benchmark traces, the MPKI reduction exceeds 2X, with the baseline MPKI in the 4-7 range (GM: 4.45) for most applications reduced to 2-3 (GM: 1.77). The improvements with an 8-way associative CDTLB follow the same trend: the baseline CDTLB MPKI GM of 4.45 drops to 1.42 with PCAX. Under fully associative designs, the CDTLB MPKI GM decreases from 2.01 (baseline) to 0.99 with PCAX.

Figure \ref{fig:iso-storage-study} combines select information presented in 
Figures \ref{fig:per-64tlb-qual-server} and \ref{fig:per-128tlb-qual-server} to allow for a direct comparison between
a 64-entry CDTLB, a 128-entry CDTLB and a 4K-entry STLB (Base-Iso), and a 64-entry CDTLB and 64-entry PCAT1 (PCAX).
The latter two take up essentially the same amount of storage, and hence we refer to this configuration as Base-Iso in the coming sections.  As we can see, PCAX outperforms the Base-Iso configuration
in reducing the MPKI in almost many cases.  Later in sections \ref{sec:CPIBenefit} and
\ref{sec:EnergyImpact} we will also see that PCAX gets better performance and consumes
less energy, respectively, than Base-Iso.
Across workloads, the GM CDTLB MPKI is 8.2 for Base-64, 4.45 for Base-Iso, and 3.4 for PCAX.

Next, we consider the impact of PCAT2 size in Figure \ref{fig:per-64tlb-qual-server-pcat2}. 
Here we have a 64-entry, 4-way set associative CDTLB and a 64-entry, 2-way set associative PCAT1.  We have plots for the effective CDTLB load MPKI for PCAT2 configurations of 512, 1K, 2K, and 4K, with 8-way set associativity in all cases, in addition to the base CDTLB (Base-64) and also simulate a configuration with no PCAT2.  PCAT2 sizes of 512 and 1K entries do quite well for many applications, but
leave some opportunity for improvement in others. A 2K entry PCAT2 is nearly as effective as the ones with 4K entries.  Referring back to Figure \ref{fig:per-static-load-server}, barring conflicts due to limited associativity, given that most applications had fewer than 2000 problem loads, we expected a 2K entry PCAT2 to be quite effective.  For many applications that had fewer than 500 problem loads, a 512-entry PCAT2 would be adequate. The results of Figure \ref{fig:per-64tlb-qual-server-pcat2} bear this out. 
Not using a PCAT2 does not achieve the benefits of PCAX in several cases. 
This is because the PCAT entry is lost when the load is evicted from the L1i and thus is not available for reuse when the load is brought back into the L1i for processing.
The GM MPKI is 8.2 for Base-64, 6.0 without PCAT2, 4.2 with a 512-entry structure, 3.6 with 1K entries, and 3.4 with both 2K and 4K entries.





\subsection {Large Page Sizes}

\begin{figure}[h]
  \includegraphics[width=1.0\linewidth]{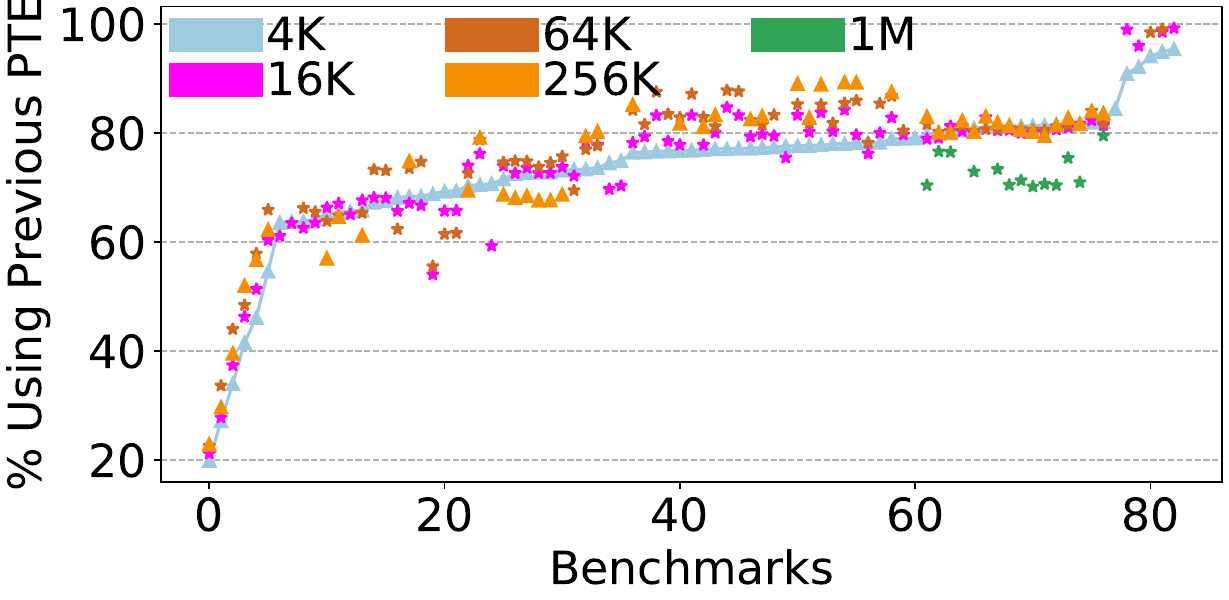}
	\caption {Large Page Study}
  \label{fig:large-page-study}
\end{figure}

\begin{figure}[t]
  \includegraphics[width=0.9\linewidth]{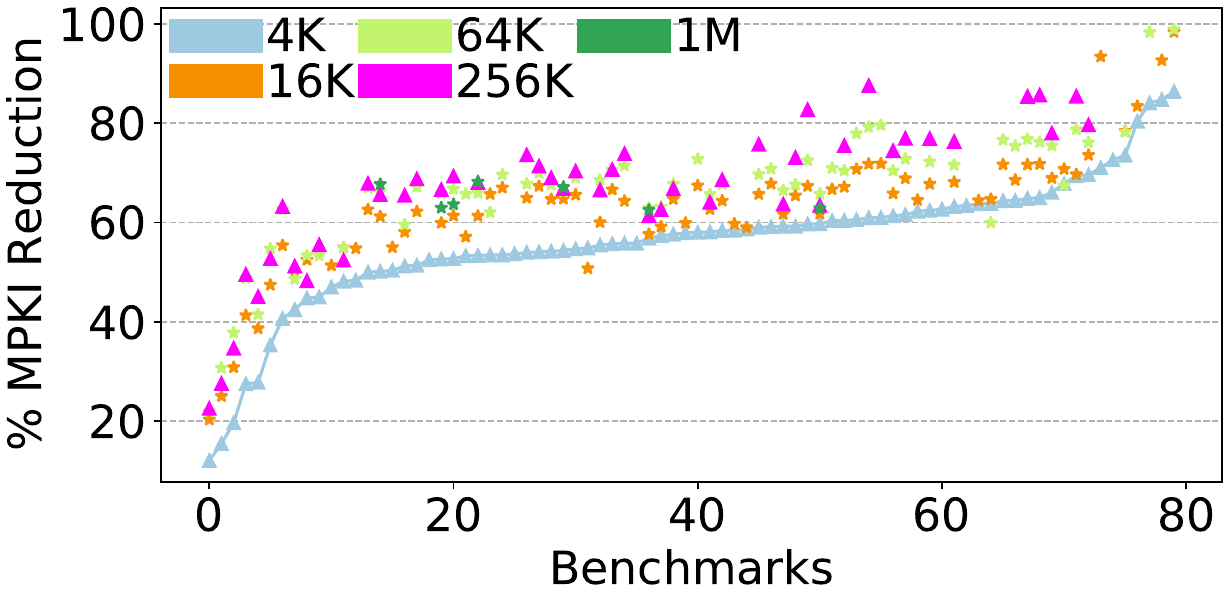}
	\caption {\% MPKI Reductions  with PCAX}
  \label{fig:large-page-imp-pcax}
\end{figure}

We now consider the effectiveness of PCAX with (uniform) page sizes larger than 4KB.  We consider only a uniform page size as our traces do not have multiple page sizes.
Figure \ref{fig:large-page-study} presents the percentage of loads that miss in a 64-entry, 4-way CDTLB that accesses the same PTE as the previous dynamic instance.
We consider page sizes of 4KB, 16KB, 64KB, 256KB, and 1MB, and present data for benchmarks where the CDTLB miss rate is greater than 1\%.
Clearly, the underlying phenomenon---that a large number of misses access the same PTE as the previous dynamic instance---is prevalent even for larger pages.
Figure \ref{fig:large-page-imp-pcax} presents the MPKI improvement for larger page sizes using PCAX. 
The CDTLB MPKI improvements continue even for larger page sizes. Note that though the MPKI reductions might be higher for larger page sizes in some cases, the base miss rates are also smaller for the large page sizes. The key takeaway is that PCAX continues to provide benefit at bigger page sizes.

\label{sec:large-page-sizes}

\begin{figure}[t]
  \includegraphics[width=0.9\linewidth]{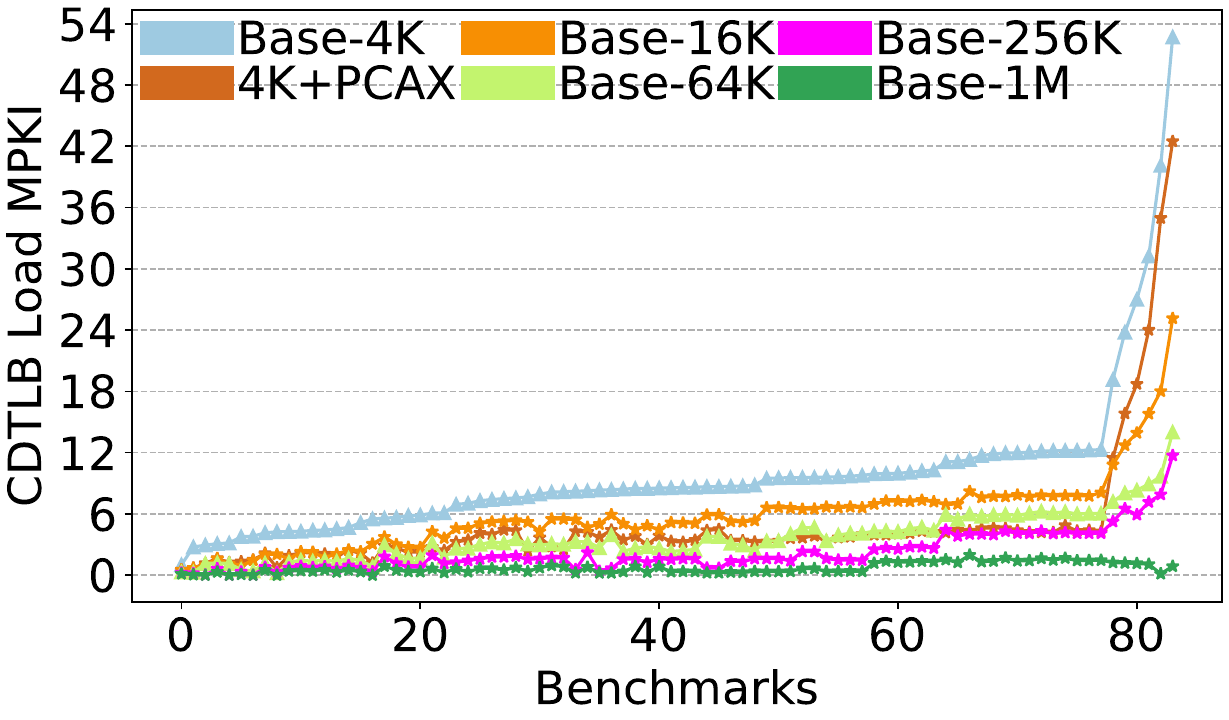}
	\caption {Large Page MPKI Comparison with 4K+PCAX}
  \label{fig:large-page-study-cmp}
\end{figure}


Figure \ref{fig:large-page-study-cmp} presents CDTLB MPKIs for large pages and also for 4KB pages with PCAX. Clearly, using larger page sizes brings down the MPKI. 
More interestingly using 4K pages with PCAX provides lower MPKIs compared to 16K and 64K page sizes. It performs as well for many benchmarks or comes close to the performance using a 256K page size.  This suggests that with PCAX larger page sizes may not be needed if a reduction in CDTLB misses is the objective.


\begin{figure*}[t]
  \includegraphics[width=1.0\linewidth]{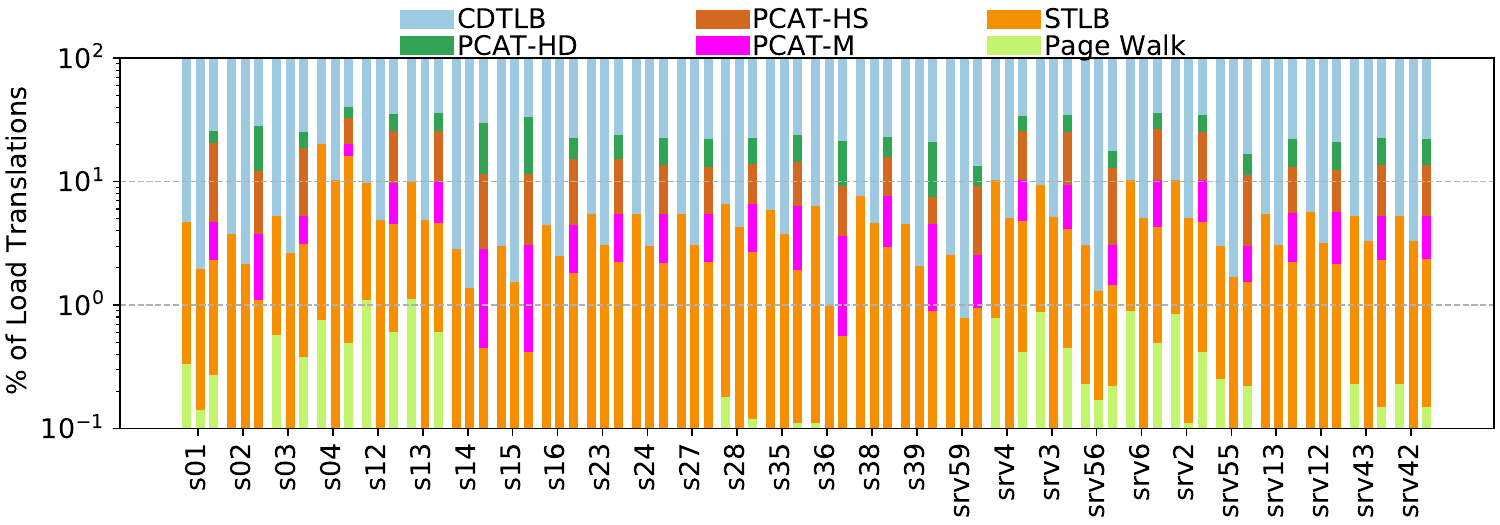}
	\caption {Data Address Translation Service Sites}
  \label{fig:pte-source-qual}
\end{figure*}

\subsection{Data Address Translation Service Sites}
\label{sec:ServiceSites}

We now consider the different sites from where a data address translation is 
serviced in Figure \ref{fig:pte-source-qual}.  Data in the figure is for a subset of the benchmarks to improve readability, as presenting data for all 84 applications would be cluttered.  Applications not considered show similar results, and we will consider all the applications when we factor this data into the CPI impact in section \ref{sec:CPIBenefit}
and energy impact in section \ref{sec:EnergyImpact}.  
For each benchmark we have three stacked bars: (i) the left bar is for a conventional 64-entry, 4-way CDTLB, 1536-entry, 12-way STLB, (ii) the middle bar is the the Base-Iso configutation (128-entry, 4-way CDTLB, 4K-entry, 16-way STLB), and (iii) the right bar is for PCAX (64-entry, 4-way CDTLB, 1536-entry, 12-way STLB, 64-entry, 2-way PCAT1, a 2K entry, 8-way PCAT2, and a 2-entry RPA buffer). 


With conventional address translation, a valid PTE can be obtained from (i) the CDTLB (labeled CDTLB), (ii) the STLB for a CDTLB miss and STLB hit (labeled STLB), or (iii) the Page Table for an STLB miss (labeled Page Walk).
With a PCAT, a valid PTE could additionally be obtained from the PCAT in three different situations: (a) the load instruction that was used to access the PCAT1 and the translation access would have missed in the CDTLB (labeled PCAT-M),
(b) the same load instruction and the translation access would have hit in the CDTLB (labeled PCAT-HS), or (c) a different load instruction whose translation access would have hit in the CDTLB (labeled PCAT-HD).  For (b) and (c) the PTE is obtained from the RPA buffer where it was placed after a PCAT1 access. Each bar for a benchmark shows the percent of load instructions that get their address translation in each of the six different ways.  The Y-axis uses a log scale so that the difference in the important portions, PCAT-M and Page Walk, can be better appreciated.

Observe that the number of references served by the CDTLB significantly in most cases because of the references served by the PCAT (PCAT-M, PCAT-HS, and PCAT-HD). There is even a non-trivial reduction in expensive Page Walks in some cases where STLB misses are non-negligible. Note that references PCAT-HS and PCAT-HD do not yield in any CPI or IPC benefit, they help mainly with energy reduction.
These reductions in CDTLB and STLB accesses, and Page Walks, lead to a lower data translation CPI and energy as we quantify next.



\subsection{CPI for Data Address Translation}
\label{sec:CPIBenefit}

In this section, we consider the improvement in the CPI for load data address translation with PCAX, borrowing from the CPI analysis in \cite{karakostas2016energy}, and the potential performance benefit.  
The parameters of the CPI model are shown in Table \ref{fig:perf-params}. Our computation of cycles spent in load address translation in conventional components consists of: (i) Cycles ($Cycles_{L1-H}$) spent in servicing CDTLB Hits ($H_{L1-H}$), where hit latency is 1 cycle (ii) Cycles ($Cycles_{L1-M}$) spent servicing CDTLB Misses ($M_{L1-M}$), which lookup the STLB that takes 8 cycles (iii) Cycles ($Cycles_{L2-M}$) spent servicing STLB misses ($M_{L2-M}$) which takes an average of 30 cycles.

\begin{figure}[t]
  \includegraphics[width=0.9\linewidth]{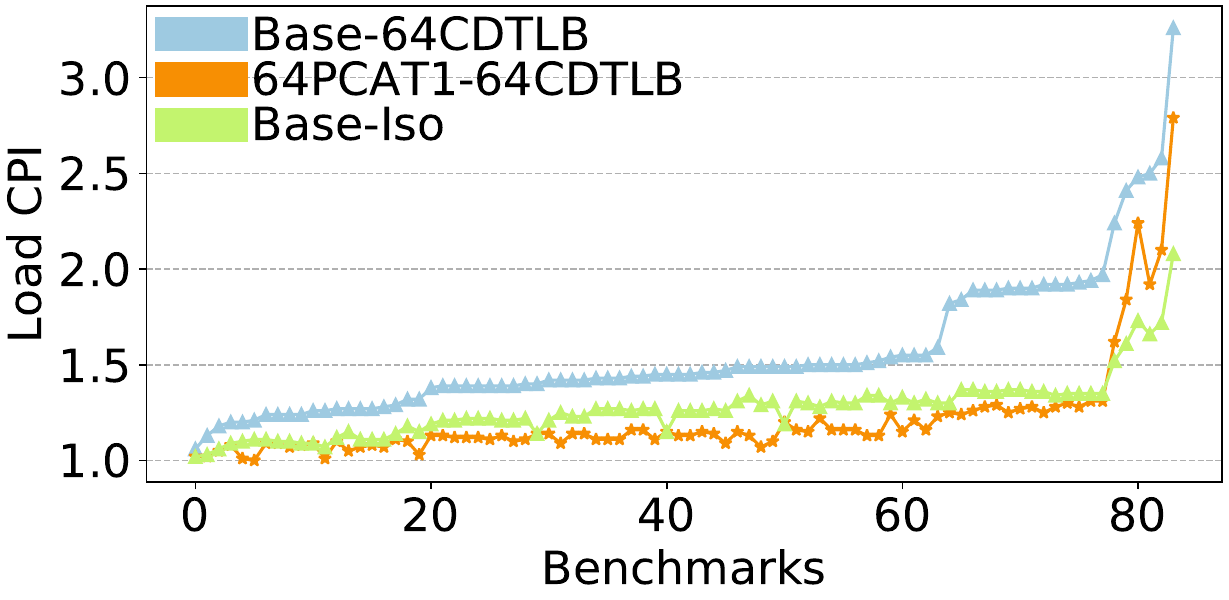}
	\caption {Data address translation CPI per load with PCAX}
  \label{fig:cpi-pcat-assist}
\end{figure}

Figure \ref{fig:cpi-pcat-assist} shows the data address translation CPI for load instructions for a 64-entry, 4-way CDTLB (Base-64), Base-Iso, and a 64-entry, 4-way CDTLB with PCAX with default PCAT1/PCAT2 sizes (PCAX) for the benchmarks.
PCAX achieves a lower  CPI, with the 64-entry CDTLB and 64-entry PCAT1 configuration having a lower CPI than even Base-Iso in most cases.
The GM CPI is 1.52 for Base-64, compared to 1.26 for Base-Iso and 1.19 for PCAX.

\begin{figure}[t]
\includegraphics[width=0.9\linewidth]{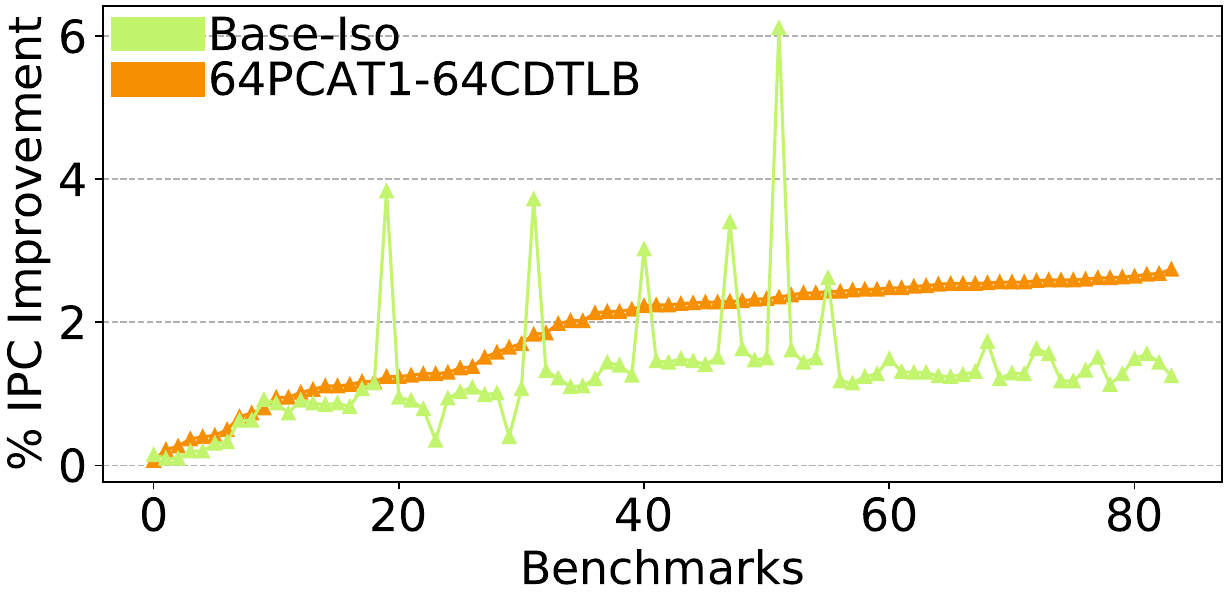}
	\caption {Performance improvement with PCAX}
  \label{fig:ipc-improv-pcat-assist}
\end{figure}

The lower data translation CPI also results in improved overall performance.
Figure \ref{fig:ipc-improv-pcat-assist} shows the percentage improvement in performance (over Base-64),
obtained from a timing simulation that incorporates a detailed TLB hierarchy and page walk modeling (as was used in \cite{vavouliotis2021exploiting,vavouliotis2021morrigan}).
\textit{Note: The benchmark ordering on the X-axis in Figure \ref{fig:ipc-improv-pcat-assist} differs from that in Figure \ref{fig:cpi-pcat-assist}.}
The performance improvement is on par with many other microarchitectural techniques. GM performance improvement with PCAX is 1.7\% over the base configuration and 1\% for the Base-Iso configuration over the base configuration.
For most benchmarks (except 6 of 84), the 64-entry CDTLB with PCAX configuration has better performance improvement than Base-Iso (even though we optimistically model the same access latency for both CDTLB/STLB sizes).


\begin{table}[]
\centering
\caption{Load Data Address Translation CPI Model}

\scalebox{0.9}{
\begin{tabular}{ll}
\hline
	\multicolumn{2}{c}{\textbf{Performance Model}}                \\ \hline
	\textbf{CDTLB Hits}                & \textit{Cycles\textsubscript{L1-H}} = 1*H\textsubscript{L1-H}                            \\
	  \textbf{CDTLB Misses}        &  \textit{Cycles\textsubscript{L1-M}} = 8*M\textsubscript{L1-M}                               \\
   \textbf{STLB Misses}        &  \textit{Cycles\textsubscript{L2-M}} = 30*M\textsubscript{L2-M}                               \\
    \textbf{Total cycles} &  \textit{Cycles\textsubscript{TLB}} = \textit{Cycles\textsubscript{L1-H}}+\textit{Cycles\textsubscript{L1-M}}+\textit{Cycles\textsubscript{L2-M}}\\\hline
     \multicolumn{2}{c}{M: Misses H: Hits}\\\hline\hline
\end{tabular}
}
\label{fig:perf-params}
\end{table}

\begin{table}[]
\centering
\caption{Dynamic Energy for Read/Write Operation}
\begin{tabular}{lllll}
\hline
\multicolumn{1}{c}{\textbf{Component}} & \multicolumn{1}{c}{\textbf{\begin{tabular}[c]{@{}c@{}}Size \\ (entr)\end{tabular}}} & \multicolumn{1}{c}{\textbf{Assoc}} & \multicolumn{1}{c}{\textbf{\begin{tabular}[c]{@{}c@{}}Read \\ (pJ)\end{tabular}}} & \multicolumn{1}{c}{\textbf{\begin{tabular}[c]{@{}c@{}}Write \\ (pJ)\end{tabular}}} \\\hline
\textbf{CDTLB}                        & 64                                                                                  & 4-way                              & 0.79                                                                              & 1.11                                                                               \\
\textbf{CDTLB-128}                        & 128                                                                                  & 4-way                              & 0.91                                                                              & 1.72                                                                               \\
\textbf{STLB}                        & 1536                                                                                & 12-way                             & 3.56                                                                              & 3.9                                                                                \\
\textbf{STLB-4K}                        & 4096                                                                                & 16-way                             &      5.61                                                                         &              8.56                                                                   \\
\textbf{PCAT1}                         & 64                                                                                  & 2-way                              & 0.25                                                                              & 0.74                                                                               \\
\textbf{PCAT2}                         & 2048                                                                                & 8-way                              & 1.3                                                                               & 3.09                                                                               \\
\textbf{L1-Cache}                      & 32KB                                                                                & 8-way                              & 9.3                                                                               & 14.31                                                                              \\\hline
\textbf{MMU-Cache-PD}                  & 32                                                                                  & 4-way                              & 0.69                                                                              & 0.9                                                                                \\
\textbf{MMU-Cache-PDP}                 & 4                                                                                   & Full                               & 0.08                                                                              & 0.02                                                                               \\
\textbf{MMU-Cache-PML4}    & 2                                                                                   & Full                               & 0.04                                                                            & 0.02                                                                             
\end{tabular}

\label{fig:energy-params}
\end{table}


\begin{figure}[t]
  \includegraphics[width=0.9\linewidth]{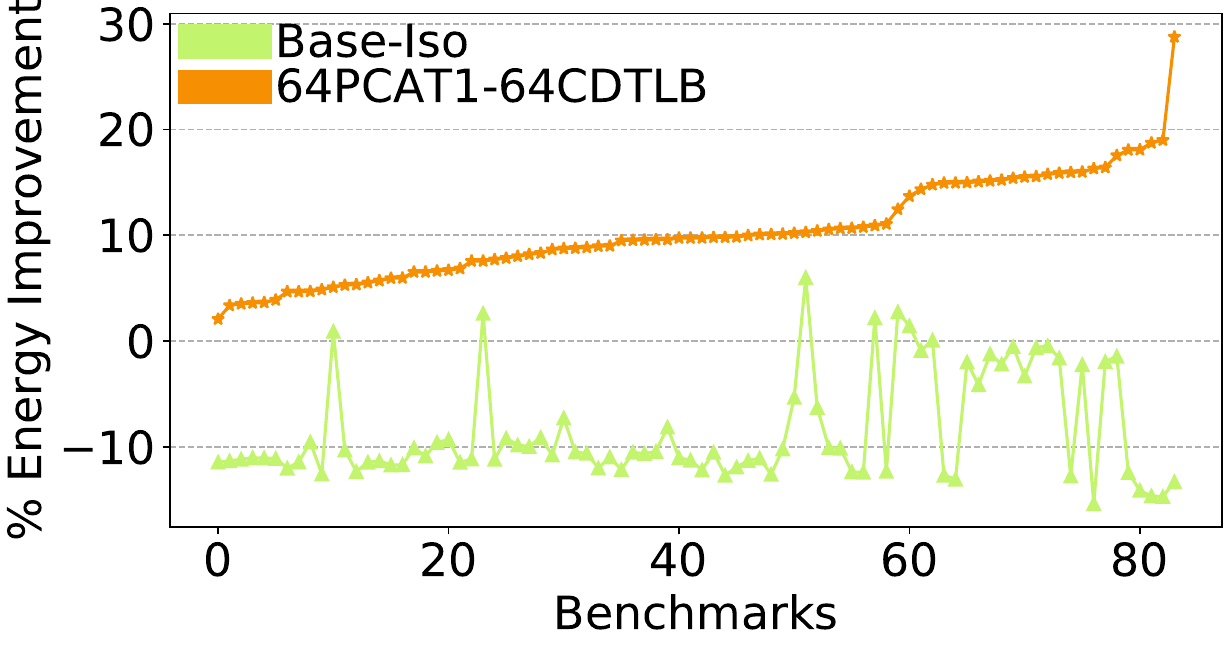}
	\caption {Energy improvement (\%) with PCAX}
  \label{fig:energy-improv-pcat-assist}
\end{figure}

\subsection{Energy Impact}
\label{sec:EnergyImpact}

Next, we consider the energy impact of PCAX over a base 64-entry, 4-way CDTLB. We measure energy consumption using CACTI 7.0 \cite{li2011cacti} with 22nm technology. The dynamic energy per Read and Write Operation for the conventional address translation structures and the PCATs are shown in Table \ref{fig:energy-params}. It is worth noting that the energies of the PCAT1 and PCAT2 are smaller than the CDTLB and STLB because they can be run a lot slower. In the case of PCAT1, it can take multiple cycles (5 in our case) to respond and still provide a translation before the address calculation.  Likewise, PCAT2 (10 cycles) is accessed alongside the (slower) L2 cache.
The longer latencies also allow for sequential tag and data access for energy savings, which we use for PCAT1 and PCAT2. The energy calculation accounts for the use of access PCAT bits, which determine the number of accesses to PCATs.


Figure \ref{fig:energy-improv-pcat-assist} shows the impact of PCAX on the dynamic data address translation energy;
\textit{the benchmark ordering on the X-axis in Figure \ref{fig:energy-improv-pcat-assist} differs from that in Figures \ref{fig:cpi-pcat-assist} and \ref{fig:ipc-improv-pcat-assist}.}
The numbers in the figure account for both loads and stores. For only loads, the energy benefit is higher as stores have little energy benefit.
There is a meaningful improvement in dynamic energy (in addition to the CPI and performance benefits seen earlier) in most cases. 
The energy reduction is higher in cases when page walks are reduced due to fewer STLB misses,
resulting in fewer L1D cache accesses due to fewer page walks, with a GM energy reduction of 7\%.
The Base-Iso configuration on the other hand, takes more energy than a 64-entry CDTLB (improvement is negative) in most cases, with a GM increase in energy consumption of 8.5\%.

\subsection {Effect of Invalidating PCATs}
\label{sec:FlushPCAT}
\begin{figure}[t]
  \includegraphics[width=0.9\linewidth]{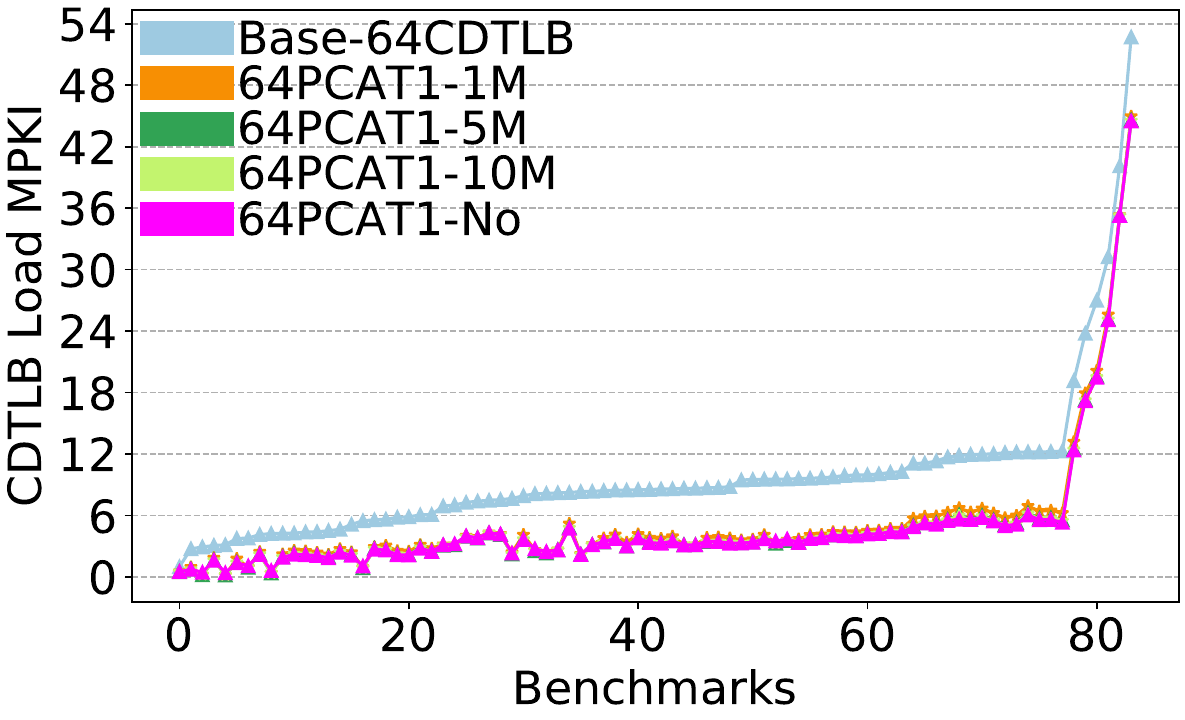}
	\caption {Impact of Periodic PCAT Invalidation}
  \label{fig:coh-overhead}
  
\end{figure}

Figure \ref{fig:coh-overhead} studies the impact of periodically invalidating the PCATs. 
Invalidating the PCATs after 5-10M instructions has almost no effect, and invalidating every 1M instruction has a marginal impact in some cases.

\subsection{Analysis on other workloads}


As mentioned above, the benchmarks we used in this paper have  have been widely used in recent papers using the Champsim simulation infrastructure, and studying a variety of issues in instruction supply, memory hierarchies, and microarchitecture.
They represent a class of server workloads with large code footprints, and in many of them data address translation is also an important problem. To assess the prevalence of the observations in section \label{sec:Motivation}, we briefly examined the translation behavior across several additional workload classes, including Google datacenter traces \cite{ranganathan2023advancing}, SPEC applications, and graph workloads.





For the Google datacenter traces for the x86 architecture, we observe similar qualitative behavior. In most traces, more than 50\% of the CDTLB misses occur when a load misses on the same PTE referenced by its previous dynamic instance. Although the overall CDTLB miss rates in these traces are generally lower than those observed in our primary workloads, approximately 50–55\% of the misses still exhibit this per-PC PTE recurrence.

For SPEC applications with CDTLB miss rates greater than 1\%, we observe strong per-PC PTE recurrence for cactuBSSN and gamess (over 90\%), and moderate recurrence for gcc (about 72\%). The effect is weaker for omnetpp and gromacs (around 25\%), and largely absent for mcf, GemsFDTD, and astar, which do not exhibit this regularity. Many other SPEC and other integer and floating point applications had a very small CDTLB miss rate.

Finally, we briefly examined graph workloads, which are often highly translation-bound due to their extremely large data footprints. Although these applications typically have a small number of static load PCs responsible for most CDTLB misses, those load PCs do not repeatedly access the same virtual page across dynamic instances (similar to mcf). Instead, their pointer-chasing behavior causes a given static load PC to reference many different pages over time, breaking the PC→PTE stability that PCAX relies upon. As a result, despite their high DTLB MPKI, graph workloads derive limited benefit from PCAX.

Overall, these observations suggest that PCAX is most effective for workloads that simultaneously exhibit: (i) non-trivial CDTLB miss rates, (ii) large instruction working sets that create temporal separation between dynamic instances of load PCs, and (iii) structured data-access patterns that preserve per-PC virtual page locality.



  

\section{Related Work}
\label{sec:Rel-work}

Given the importance of data address translation, a large body of work aims to reduce DTLB misses, i.e., DTLB accesses that fall back to secondary TLBs or page tables. Excellent overviews appear in recent papers \cite{gosakan2023mosaic,vavouliotis2021exploiting,guvenilir2020tailored,karakostas2016energy} and we provide a brief overview of some of the prior work below.

A significant line of work proposes novel DTLB designs. Several designs exploit virtual and physical page contiguity to reduce the number of TLB entries and misses \cite{pham2012colt,pham2014increasing}. Hardware and OS support for large pages \cite{thp} similarly improves TLB reach, and many designs tailor the TLB to better support huge pages \cite{cox2017efficient,papadopoulou2015prediction,karakostas2016energy}. Since contiguity may be limited by fragmentation, techniques have also been proposed to mitigate this challenge \cite{basu2013efficient,park2017hybrid,karakostas2015redundant}. Beyond discrete large-page sizes, \cite{guvenilir2020tailored} exploits intermediate levels of contiguity. Other recent work reorganizes page tables and employs hashing to significantly reduce miss rates \cite{gosakan2023mosaic}. Several proposals cache page-table components (e.g., via MMU and PTE caches) to reduce TLB misses and page-walk latency \cite{barr2010translation,bhattacharjee2013large,ryoo2017rethinking,intelpw}.

Another thread of work reduces misses through prefetching—primarily into secondary TLBs—by exploiting memory access stream patterns \cite{bhattacharjee2009characterizing,kandiraju2002going,saulsbury2000recency}. Recent work combines multiple prefetchers and selects between them based on usefulness \cite{vavouliotis2021exploiting}, also leveraging locality in the page table to prefetch nearby PTEs.

PC‑based TLB prefetchers (\cite{kandiraju2002going,vavouliotis2021morrigan}) appear superficially related to PCAX because both use the program counter (PC) as a correlating signal. However, their goals, mechanisms, and semantics differ fundamentally. PC‑based prefetchers use the PC to predict future virtual page numbers and speculatively prefetch those pages into the STLB. The PC serves only as a trigger to initiate speculative requests. PCAX does not predict future VPNs: instead, it stores and supplies the actual PTE most recently missed by that static load PC. Rather than prefetching into the TLB hierarchy, PCAX provides an alternative translation path, bypassing the CDTLB entirely for select loads by returning a PTE at fetch time. As a result, PCAX is not a prefetching mechanism; it is a PC‑driven address‑translation path that directly supplies the PTE when the PC reappears.

In contrast to memory‑address‑stream speculation, PCAX exploits the stability of translation behavior for specific load PCs, binding translations directly to instruction identity rather than virtual‑address patterns. This enables PCAX to access the PTE during instruction fetch, well before the data address is computed or the CDTLB is accessed. Because the PTE is not needed until the address‑generation stage, PCAX can perform its lookup slowly and energy‑efficiently without impacting performance. This early, PC‑driven mechanism provides PTEs for select loads ahead of time, improving both performance and data‑translation energy, and operates orthogonally to both STLB replacement and prior prefetching techniques.


The VAX processors used the concept of \textit{pre-translation}, which observed that multiple dynamic instances of instruction fetching accessed the same PTE and optimized it by maintaining the PTE in a special register \cite{satyanarayanan1981design}. In a similar vein, our design incorporates an RPA buffer that opportunistically reuses recently supplied translations from PCAX.

Regarding optimizations to the processing of load instructions based upon their identity rather than data address,
researchers have also made empirical observations that a small number of static load instructions account for a disproportionate number of data cache misses \cite{abraham1993predictability,tyson1997managing}.

\section{Concluding Remarks}
\label{sec:Conclusions}

We presented a novel technique that uses the PC of a load instruction, rather than the address of the
referenced data, to obtain an PTE for the referenced data.
The key observation driving the use of a PC rather than the data address is that often
a dynamic instance of a load instruction accesses the same virtual page
as the prior dynamic instance, thus needing the same PTE.
This technique can only be profitably used by a small subset of the static loads in the program.
We further observed that only a small subset of the static loads accounts for most of the
address translation misses with conventional DTLBs.
Combining these two observations we proposed \textit{PC-Indexed Data Address Translation (PCAX)},
to assist conventional address translation structures.
PCAX employs tables that require additional storage, which is modest given the transistor
budgets in contemporary high-performance chips, but is able to achieve a significant reduction (2-3X or more) in the
effective DTLB miss rate, and improve performance, 
while also reducing the energy for data address translation in most of the considered benchmarks. PCAX is a purely microarchitectural solution that is highly effective, \textit{with no demands on the ISA, software and/or systems}, unlike other proposed techniques which have varying consequences for system software.

Future work includes assessing if the observations that underlie PCAX can be applied to create new prefetching or replacement schemes for an STLB, and perhaps even a CDTLB, with or without PCAX, to further improve translation efficiency.




\bibliographystyle{IEEEtranS}
\bibliography{refs}

\end{document}